\documentclass[12pt]{article}
\usepackage{amssymb,graphicx,epsfig} 
\def\etal{{\em et al.}}
\def\issue(#1,#2,#3){{\bf #1}, #2 (#3)} 

\def\APP(#1,#2,#3){{\it Acta Phys.\ Polon.} \ \issue({\bf #1},#2,#3)}
\def\ARNPS(#1,#2,#3){{\it Ann.\ Rev.\ Nucl.\ Part.\ Sci.} \ \issue({\bf #1},#2,#3)}
\def\CPC(#1,#2,#3){{\it Comp.\ Phys.\ Comm.} \ \issue({\bf #1},#2,#3)}
\def\CIP(#1,#2,#3){{\it Comput.\ Phys.} \ \issue({\bf #1},#2,#3)}
\def\EPJ(#1,#2,#3){{\it Eur.\ Phys.\ J.} \ \issue({\bf #1},#2,#3)}
\def\EPJD(#1,#2,#3){Eur.\ Phys.\ J. Direct\ C \ \issue({\bf #1},#2,#3)}
\def\IEEETNS(#1,#2,#3){{\it IEEE Trans.\ Nucl.\ Sci.} \ \issue({\bf #1},#2,#3)}
\def\IJMP(#1,#2,#3){{\it Int.\ J.\ Mod.\ Phys.} \ \issue({\bf #1},#2,#3)}
\def\JHEP(#1,#2,#3){{\it J.\ High Energy Physics} \ \issue({\bf #1},#2,#3)}
\def\JP(#1,#2,#3){{\it J.\ Phys.} \ \issue({\bf #1},#2,#3)}
\def\MPL(#1,#2,#3){{\it Mod.\ Phys.\ Lett.} \ \issue({\bf #1},#2,#3)}
\def\NP(#1,#2,#3){{\it Nucl.\ Phys.} \ \issue({\bf #1},#2,#3)}
\def\NIM(#1,#2,#3){{\it Nucl.\ Instrum.\ Meth.} \ \issue({\bf #1},#2,#3)}
\def\PL(#1,#2,#3){{\it Phys.\ Lett.} \ \issue({\bf #1},#2,#3)}
\def\PR(#1,#2,#3){{\it Phys.\ Rev.} \ \issue({\bf #1},#2,#3)}
\def\PRL(#1,#2,#3){{\it Phys.\ Rev.\ Lett.} \ \issue({\bf #1},#2,#3)}
\def\SJNP(#1,#2,#3){{\it Sov.\ J. Nucl.\ Phys.} \ \issue({\bf #1},#2,#3)}
\def\ZP(#1,#2,#3){{\it Zeit.\ Phys.} \ \issue({\bf #1},#2,#3)}
\def\be {\begin{equation}}
\def\ee {\end{equation}}
\def\bea {\begin{eqnarray}}
\def\eea {\end{eqnarray}}
\begin{document}
\begin{titlepage}
\begin{flushright}
CU-PHYSICS/07-2009 \hspace*{1.3in} HIP-2009-26/TH \hfill TIFR/TH/09-37
\end{flushright}
\begin{center}
{\LARGE {\bf
{Multijet Discriminators for New Physics  \\ [2 mm]
in Leptonic Signals at the LHC}}} \\[5mm]
\bigskip
{\sf Biplob Bhattacherjee} $^a$, 
{\sf Anirban Kundu} $^b$, 
{\sf Santosh Kumar Rai} $^c$ and
{\sf Sreerup Raychaudhuri} $^a$

\bigskip\bigskip

$^a${\footnotesize\rm 
Department of Theoretical Physics, Tata Institute of Fundamental Research, \\ 
1, Homi Bhabha Road, Mumbai 400 005, India. \\
E-mail: {\sf biplob@theory.tifr.res.in, \sf sreerup@theory.tifr.res.in} }

\bigskip
$^b${\footnotesize\rm 
Department of Physics, University of Calcutta, \\
92, Acharya Prafulla Chandra Road, Kolkata 700 009, India. \\
E-mail: {akphy@caluniv.ac.in} }

\bigskip
$^c${\footnotesize\rm 
Department of Physics, University of Helsinki and Helsinki Institute of 
Physics, \\ P.O. Box 64, FIN-00 014, University of Helsinki, Finland. \\
E-mail: {\sf santosh.rai@helsinki.fi} }

\normalsize
\vskip 10pt

{\large\bf ABSTRACT}
\end{center}

\begin{quotation} \noindent\small 
Some of the cleanest signals for new physics in the early runs of the 
LHC will involve strongly-produced particles which give rise to multiple 
leptons by undergoing cascade decays through weakly-interacting states 
to stable particles. Some of the most spectacular final states will 
involve three or more leptons, multiple jets and generally missing 
energy-momentum as well. A triad of the most interesting models of new 
physics which induce such signals is known to consist of ($i$) 
supersymmetry with $R$-parity conservation, ($ii$) a universal extra 
dimension with conservation of KK-parity and ($iii$) little Higgs models 
with conserved $T$-parity. Similar signals could also arise if the 
Standard Model is augmented with a fourth sequential generation of heavy 
fermions. We study all these possibilities and show that a judiciously 
chosen set of observables, critically involving the number of 
identifiable jets and leptons, can collectively provide distinct 
footprints for each of these models. In fact, simple pairwise 
correlation of such observables can enable unambiguous identification of 
the underlying model, even with a relatively small data sample.
\vskip 10pt
\normalsize
PACS numbers: {\tt 12.60.Jv, 14.80.Rt, 12.60.Fr} \\
\end{quotation}
\begin{flushleft}\today\end{flushleft}
\vfill
\end{titlepage}
\newpage
\setcounter{page}{1}

\section{Introduction}

\noindent After some initial glitches, the Large Hadron Collider (LHC) 
at CERN, Geneva, is expected to start colliding protons with protons at 
the end of the current year (2009). Once the calibration stage is over, 
the start-up centre-of-mass energy has been fixed at 7~TeV and it is 
later expected to increase through 10~TeV to the final goal of 14~TeV 
\cite{LHCwebsite}. Initial luminosity targets are somewhat modest -- 
being estimated at 60~pb$^{-1}$ in the first year. Though no official 
statements on further progress are available, educated guesses put it at 
around 1~fb$^{-1}$ in about two years of running, and perhaps 
5~fb$^{-1}$ at the end of three years. Thus, an eventual goal of 30 (or 
even 50~fb$^{-1}$) integrated luminosity seems not unattainable, 
assuming that the LHC will run without unscheduled breaks for at least 
10 years, with some later upgrade in luminosity. At the present 
juncture, 100~fb$^{-1}$ of integrated luminosity seems definitely 
unattainable.

\bigskip\noindent The search for signals of new physics at the LHC, 
especially in the early runs, will be based on three requirements, viz. 
large cross-sections, clean triggers and distinctive final states. 
Obviously, large cross-sections can be obtained if the underlying 
process involves strong interactions. A very rough estimate of the 
cross-section for a $2 \to 2$ process of this kind is
\begin{equation} 
\sigma \approx \frac{\alpha_s(s)^2}{s} 
\end{equation} which, for 
$\sqrt{s} \approx 1$~TeV, comes out as $\sigma \approx 4$~pb. With the 
luminosity estimates mentioned above, this would mean 4~000, 20~000, and 
at least 120~000 events at the end of 2, 3 and 10 years respectively. 
Thus, we shall have fairly copious production of the particles in 
question.

\bigskip\noindent Particles which are produced through strong 
interactions will generally decay through strong interactions, unless 
forbidden to do so by some conservation law. One would, therefore, 
expect the principal decay modes following production of new, 
strongly-interacting particles to produce hadronic final states, 
generically with a small number of identifiable jets. Such signals would 
probably be completely lost in the huge QCD backgrounds expected at a 
high energy proton-proton machine. For example, the cross-section for 
dijet production at the LHC is of the order of microbarns 
\cite{LHCdijetCS}, which implies a few million dijet events in the first 
two years of running. For this reason, in the messy environment of a 
hadron machine, clean triggers are traditionally associated with {\it 
leptonic} final states -- with or without large amounts of missing 
energy and momentum. This allows us to classify signals obtainable at 
the LHC into three types, viz.

\begin{enumerate}

\item Final states containing easily-tagged leptons which are produced 
directly through interactions of electroweak strength.

\item Final states containing easily-tagged leptons which are produced 
in cascade decays of parent particles which are produced through strong 
interactions.

\item Final states containing hadronic jets which arise either directly 
from strong interactions or from hadronic decays of parent particles 
produced in strong interactions.

\end{enumerate}
The first two types of signals are easy to tag and study, and the last 
two correspond to large cross-sections. This means that the most 
promising signals correspond to the second type, i.e. leptonic final 
states arising from cascade decays of strongly produced particles. These 
have large cross-sections and are easy to tag as well. However, in order 
that such signals should occur,

\begin{itemize}

\item the `new physics' must have a very definite kind of mass spectrum, 
with two classes of particles, viz. ($i$) the strongly-interacting `new' 
particles, which should be heavier than the ($ii$) weakly-interacting 
`new' particles which decay to leptons, and

\item there should be a conserved quantum number (or nearly-conserved 
quantum number) carried by all the `new' particles which ensures that 
the first class of particle does not decay directly to 
strongly-interacting Standard Model (SM) particles, but cascades down 
through particles of the second class.

\end{itemize}

\noindent These conditions automatically restrict us to a small set of 
existing models, of which only {\it four} may be considered popular 
options. These are:
\begin{enumerate}

\item {\it Supersymmetric models with conservation of $R$-parity}: It 
has been known for a long time that in the constrained minimal 
supersymmetric model (cMSSM), which is based on minimal supergravity 
(mSUGRA) and universality of parameters at a very high 
scale\cite{cMSSM}, the strongly interacting sparticles -- the squarks 
and gluinos -- though degenerate with the weakly-interacting ones at the 
gauge unification scale, generally tend to become heavier at the TeV 
scale because of the renormalisation group (RG) evolution of their 
masses. These squarks and gluinos undergo cascade decays down to the 
lightest supersymmetric particle (LSP), which is stable and invisible if 
$R$-parity is conserved, producing leptons on the way. Extensive studies 
of such signals are available in the literature\cite{SUSYatLHC}.

\item {\it Universal extra dimension models with conservation of 
$KK$-parity}: In the UED(5) model, which has an extra spatial dimension 
with the topology of a circle folded about one of its diameters, i.e. 
$\mathbb{S}^{(1)}/\mathbb{Z}_2$, the extra dimension can be accessed by 
all the SM fields\cite{UED5model}. The Kaluza-Klein (KK) excitations of 
all fields which belong to the same Kaluza-Klein number ($n = 1, 2, 
\dots$) are degenerate at the tree-level, but this degeneracy is split 
by radiative corrections\cite{UED5spectrum}. Stronger interactions and 
larger colour factors then combine to make the KK excitations of quarks 
($q_n$) and gluons ($g_n$) heavier than the others. These 
strongly-interacting particles then undergo cascade decays down to the 
lightest KK particle (LKP), which is stable and invisible if KK-parity 
$(-1)^n$ is conserved, producing leptons on the way. Such signals have 
also been studied\cite{UED5signals}, though not with the same level of 
detail and sophistication as the corresponding signals from 
supersymmetry.

\item {\it Little Higgs models with conservation of $T$-parity}: Little 
Higgs models have an extended gauge symmetry which is spontaneously 
broken, yielding a light Higgs boson as a Nambu-Goldstone boson, which 
acquires its mass through small radiative effects. The presence of two 
sets of gauge bosons and scalars, forming irreducible representations of 
two direct product gauge groups, ensures that the dominant terms in the 
radiative corrections to the light Higgs boson cancel, thereby 
postponing the hierarchy problem to a scale of some tens of TeV, which 
is out of the kinematic reach of the LHC\cite{LittleHiggs}. 
Phenomenological constraints from electroweak precision tests require 
the introduction of a conserved quantum number called $T$-parity, which 
protects the other SM particles from acquiring unacceptably large 
masses\cite{LHTmodel}. The LHC signals of this LH(T) model have been 
studied\cite{LHTsignals}, but nowhere near as comprehensively as in the 
previous two cases. In fact, this model contains heavy $T$-odd partners 
of the quarks, which can undergo cascade decays down to the lightest 
$T$-odd particle (LTP), which is stable and invisible if $T$-parity is 
conserved, producing leptons on the way. The requirement that the stable 
LTP, a dark matter candidate, be neutral and weakly-interacting, ensures 
that the $T$-odd quarks will be heavier than {\it at least} one other 
particle, which, in turn, ensures that there may be three-body decays 
with leptons, if not cascades.

\item {\it A fourth sequential generation with heavy $b_4$ quarks}: 
Though precision electroweak tests rule out the presence of a SM with a 
degenerate fourth generation, it is still possible\cite{SM4facts}, with 
a modest amount of fine-tuning, to accommodate a heavy fourth generation 
with non-degenerate SU(2)$_L$ partners. All that is needed is a mass 
pattern such that the fourth generation contribution to the $S$ 
parameter
\begin{equation} 
S_4 = \frac{2}{3\pi} - \frac{1}{3\pi}\left[\log\frac{m_{t_4}}{m_{b_4}} - 
\log\frac{m_{\nu_4}}{m_{\tau_4}} \right]
\label{eqn:Sparameter_in_SM4}
\end{equation} 
lies within the experimental error. With four unknown parameters, viz. 
$m_{t_4}, m_{b_4}, m_{\nu_4}$ and $m_{\tau_4}$, this is not difficult to 
arrange. At the same time, if $|m_{t_4} - m_{b_4}|$ is greater than 
about 50 GeV, one gets too large a contribution to the $T$ 
parameter\cite{SM4facts}. Given such a spectrum, and reasonable values 
of the 4$\times$4 Cabibbo-Kobayashi-Maskawa (CKM) matrix elements 
$V_{t_4b} \approx V_{tb_4} \sim 0.1$, it is possible for the $t_4$ and 
$b_4$ to decay to final states containing two or more leptons. In this 
case, the approximate conservation of flavour implied by the smallness 
of the off-diagonal CKM elements ensures that there will be cascade 
decays, not through new particle states, but through heavy flavour 
states (mostly $t$ and $b$ quarks), with leptons as the end product. 
This SM(4) scenario, the simplest of all four options, produces 
similar-looking signals to the other cases, even though it does not 
offer a dark matter candidate. 
\end{enumerate}

\noindent Of the above models, the first three, viz. cMSSM, UED(5) and 
LH(T) have some attractive features in common. The Higgs boson mass is 
protected from large radiative corrections, either by cancellation, or 
by bringing in new physics at a scale of a few TeV to a few tens of TeV, 
and there exists a stable, weakly-interacting particle (the LSP, LKP or 
LTP), which is an excellent dark matter candidate. The overwhelming 
observational evidence that dark matter exists in the cosmos and is 
non-baryonic in nature\cite{DarkMatter}, has made it a real challenge to 
find out its nature. Thus, models like the above, which not only solve 
the decades-old hierarchy problem but also provide a simple theory of 
dark matter, lie at the forefront of the hypotheses that we would like 
to test when the LHC begins to probe the TeV scale. At the same time, a 
fourth generation is a simple and natural extension of the SM, and may 
offer clues to the mystery of the pattern of fermion masses and mixings. 
Hence, although the SM(4) scenario neither proffers a dark matter 
candidate nor prevents the Higgs boson from picking up large self-energy 
corrections, its very simplicity would make it a natural theme of study 
at the LHC.

\bigskip \noindent Having established that all of the four models 
discussed above are interesting in their own right and can yield 
similar-looking multi-lepton event topologies at the LHC, the question 
immediately arises: {\it If some excess over the SM prediction is indeed 
observed at the LHC in the multi-lepton channels, which of these models 
could be the likely explanation?} This question is a specific case of 
the larger `LHC inverse problem', which addresses the general issue of 
the nature and properties of any new physics which may be discovered at 
the LHC. Experience from the past tells us that if indeed some deviation 
from the SM is announced, it is sure to result in a scramble among 
particle theorists to prove how this deviation fits in with one's 
favourite model! Quite likely, {\it more} than one of the above models 
will fit the observed events for some specific choice -- not necessarily 
unique -- of the model parameters. However, again experience from the 
past tells us that the hypothesis which fits one particular set of data 
-- say the invariant mass distribution -- may not fit a different set of 
data -- say the angular distribution -- for the same set of events. Only 
if one chooses the correct model should we expect to get a good fit 
across all kinds of observables. It would be like the case of the shoe 
fitting all the available footprints and leading to an unambiguous 
identification. However, except for some exploratory 
attempts\cite{LHCinverse}, the LHC inverse problem has not been studied 
in full seriousness by the high energy physics community. There has been 
an attitude of watch-and-wait with most workers in the field, expecting 
to actually see a deviation from the SM in the LHC runs and only then to 
proceed to fit it, as was the practice in the past. Against this, it is 
argued that the LHC and its physics calls for a new approach. Given the 
level of expectation aroused by this machine, the exciting fact that it 
explores a hitherto-untouched energy regime and the more sobering fact 
that there is no obvious successor of its kind, the smallest hint of new 
physics at the LHC would call for an immediate theoretical analysis, 
with the inverse problem following hard on the heels of any announced 
deviation.

\bigskip\noindent Can we devise a quicker method to identify new physics 
than the traditional $\chi^2$-fitting (which would, of course, have to 
be done eventually)? To be successful, such a method would have to be 
simple and robust and preferably economic in the sense that it should 
use data which will be collected anyway. Keeping this in mind, the 
present study focusses on final states at the LHC which have \\[1mm]
\hspace*{0.3in} ($a$) \ 3 or 4 {\it hard leptons}, i.e. $3\ell$ and 
$4\ell$; \\[1mm]
\hspace*{0.3in} ($b$) \ substantial {\it missing energy}~ 
$/\!\!\!\!E_T$, and \\[1mm]
\hspace*{0.3in} ($c$) \ an indeterminate number of {\it hadronic jets}, 
i.e. $nJ$ where $n = 0, 1, 2, 3, \dots$. \\[1mm]
Such states would stand out among the LHC events and would be among the 
first where signals for new physics could be sought. SM backgrounds to 
such states arise principally from $WZ$ or $ZZ$ production, which are 
necessarily electroweak and hence would have a cross-section of the 
order of a few hundred femtobarns. Given the small branching ratios of 
$W$ and $Z$ to leptons which can be tagged at the LHC, i.e. electrons 
($e^\pm$) and muons ($\mu^\pm$), this means that the background is at 
the level of only a few events, if at all, for around 30~fb$^{-1}$ worth 
of data. In the kind of new physics models described above, the 
probability of obtaining the same final states is one -- or even two -- 
orders of magnitude greater, which means that if something like a few 
hundred of such events is seen, we shall have a `smoking gun' signal for 
new physics beyond the SM. These $3\ell$ and $4\ell$ signals are, in 
fact, much better hunting grounds for new physics than signals with one 
or two final state leptons, because the latter have large irreducible SM 
backgrounds from resonant $W$ and $Z$ production as well as $t\bar{t}$
production.

\bigskip\noindent Leptonic signals at the LHC with missing energy and 
with or without hadronic jets, have, of course, been studied and 
described several times in the literature -- particularly in the context 
of SUSY searches\cite{SUSYleptons} -- and it is well known that they 
will stand out from the SM background rather conspicuously. It is not 
our purpose, in this work, to reinvent this particular wheel. Rather, 
using the information that these are indeed viable new physics signals, 
we seek to illustrate how the four models described above can lead to 
similar-looking signals, and then to consider ways and means for {\it 
quickly distinguishing} between these models using the experimental data 
which is likely to be available. We reason that each model {\it must} 
leave its distinctive imprint in the final states, in the form of 
kinematics reflecting the mass spectrum, branching ratio structures and 
angular distributions, as well as jet and lepton multiplicities-- and 
our problem is to identify and dig out the correct clues. This is the 
motivation and principal theme of the present work.

\bigskip\noindent This article is organised as follows. In Section 2, we 
discuss in more detail how ($3\ell$ + MET + $n$J) and ($4\ell$ + MET + 
$n$J) signals can arise in each of the above models, and what would be 
the distinctive characteristics of events arising from each model. In 
Section 3, we explain our construction of discriminating variables, 
followed by some educated guesswork on the kind of values predicted for 
the discriminating variables. Section 4 mentions some essential features 
of our numerical analysis and describes our results. Finally, Section 5 
comprises a summary and critique of our work.

\section{Comparative Anatomy of Multi-Lepton Signals}

In the previous section, we had made a very rough estimate that the 
cross-section for the pair production of two strongly-interacting new 
particles with masses in the range of a few hundred GeV to a TeV would 
be of the order of a few picobarns. However, even assuming that the 
coupling strengths resemble QCD couplings, the real cross-sections will 
actually depend on several factors, such as ($i$) the number of new 
particles being produced, ($ii$) the exact masses of these particles, 
($iii$) the number of Feynman diagrams contributing to the process and 
possible interference effects, ($iv$) the actual machine energy and 
last, but not least, ($v$) the parton density functions (PDFs) being 
used for the prediction, especially the scale $Q^2$ at which the PDFs
are being evaluated. Apart from this, there will be sub-leading 
effects like initial and final-state radiative corrections, 
multiparticle interactions as well as detector issues such as energy and 
angular resolution and irreducible effects of pileup etc. In the present 
study, all these minor effects have been neglected and we take only the 
leading order (LO) cross-sections convoluted with PDFs -- also at LO. A 
justification for this will be given in the next section. The 
calculations have been carried out using the event generator 
PYTHIA\cite{pythia}, cross-checked and/or interfaced with the CalcHEP 
software\cite{calchep}. For the PDF's, we have used the CTEQ-5 
sets\cite{CTEQ5} inbuilt in PYTHIA. The advantage of using PYTHIA, apart 
from the fact that it is widely used by theorists and experimentalists 
alike, is that fragmentation and hadronization of partons is taken care 
of, and a rudimentary jet-formation algorithm can be easily implemented 
using the inbuilt PYCELL routine. More details will be presented in the 
context of individual models of new physics.

\bigskip\noindent Far more significant than the subleading corrections 
to the LO calculation of the cross-section is the fact that there is 
still some uncertainty as to the actual machine energy at which the 
experimental data will be collected. It appears to be quite reasonable 
to hope that this will soon rise above the tentative start-up value of 
7~TeV, but caution must be exercised in assuming that the energy will 
thereupon be speedily upgraded to the eventual goal of 14~TeV. This may, 
in fact, take several years, and hence, an intermediate energy value of 
10~TeV seems to be the favoured choice for LHC studies at the present 
juncture. This has been followed in the present work. It is, however, 
entirely possible that the early performance of the LHC will be so 
satisfactory that the energy does gets upgraded to 14~TeV quite soon. In 
that case, we have checked that there will be little {\it qualitative} 
difference in our results, though actual numbers will, naturally, be 
different.

\begin{figure}[htb]
\setcounter{figure}{0}
\centerline{ \epsfxsize= 6.5 in \epsfysize= 2.8 in \epsfbox{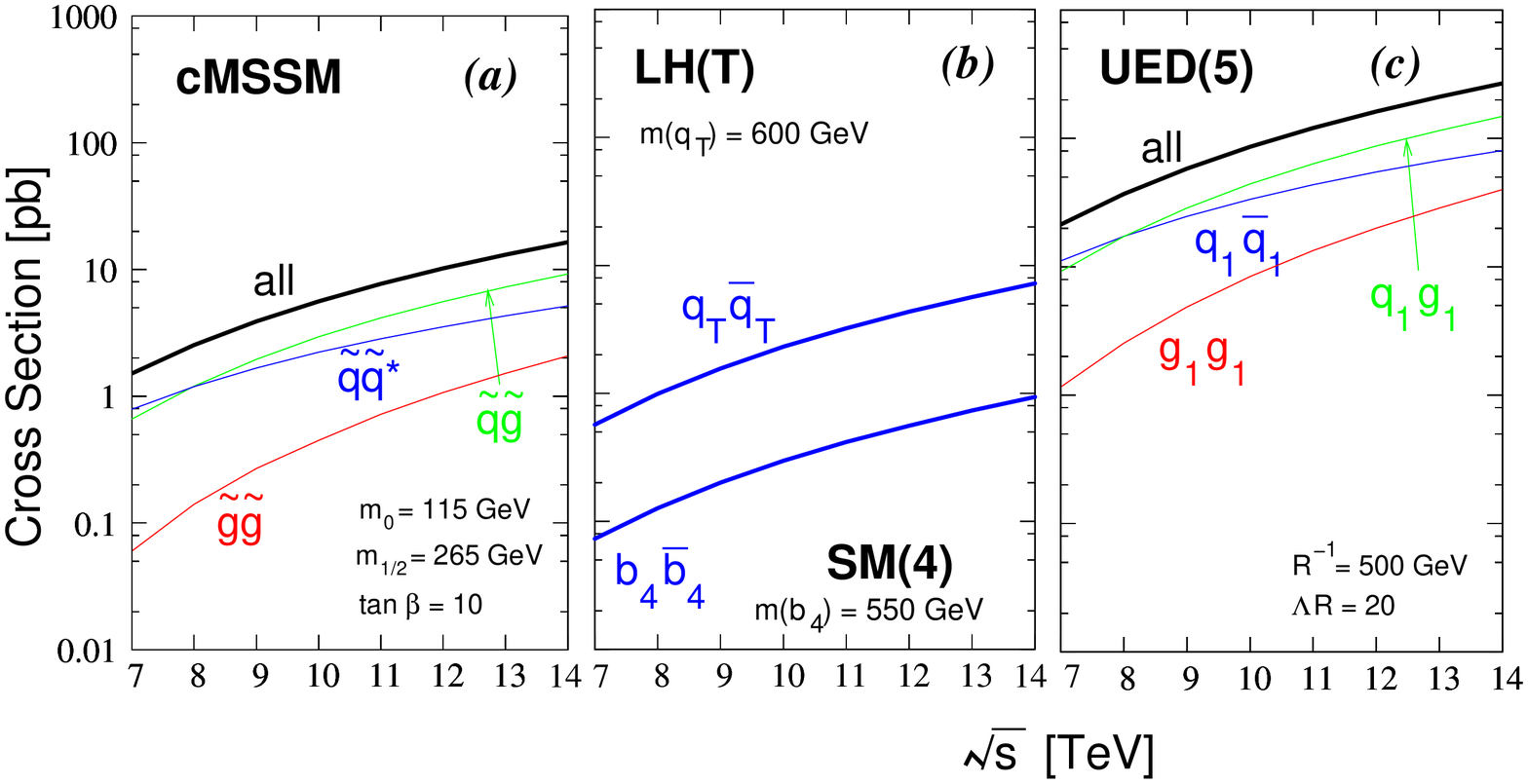} }
\vskip -10pt
\caption{{\footnotesize\it Cross-sections for strong pair-production of 
new particles at the LHC as a function of the machine energy. The thick 
lines represent the total cross-section in each model. Plotted in the 
boxes, as marked, are ($a$) the {\em cMSSM} with $A_0 = 0$, $\mu > 0$, 
and other parameters as marked, ($b$) the {\em LH(T)} model with $f = 
1$~TeV and the {\em SM(4)} with $m(b_4) = 550$~GeV, and ($c$) the {\em 
UED(5)} model. Note that $q$ stands generically for all quarks and their 
heavy counterparts.}}
\label{fig:CrossSections}
\end{figure}
\vskip -10pt

\bigskip\noindent A rough understanding of how our numerical results 
will change as the machine energy varies from 7~TeV through 10~TeV to 
14~TeV can be obtained by inspecting the graphs in 
Figure~\ref{fig:CrossSections}. These show, for an illustrative choice 
of a single point in the parameter space of each of the four models 
mentioned in the last section, the variation of the cross-section for 
the production of the main strongly-interacting particles as the machine 
energy $\sqrt{s}$ varies from 7 -- 14~TeV. The box on the left, marked 
($a$), shows the supersymmetric cross-sections, with squark production, 
in association with an anti-squark or a gluino, dominating the gluino 
pair production. The squark and the gluino are around 600~GeV and 
650~GeV, respectively, in this plot. This makes them much lighter than 
the machine energy and indicates that a large part of the cross-section 
comes from initial-state partons with very small momentum fractions. The 
steep rise in PDFs, especially gluon PDFs, for small arguments, is 
well-known and accounts for the fact that these cross-sections, despite 
being proportional to $\alpha_s^2/\hat{s}$, actually grow with energy 
$\sqrt{s}$.

\bigskip \noindent The central box, marked ($b$) shows cross-sections 
for the pair production of heavy quarks in the LH(T) and SM(4) models. 
In the LH(T) model the cross-section is for the heavy $T$-odd partners 
of the generic quark $q$, which are taken to have a common mass around 
600~GeV, while in the SM(4), the isospin $-\frac{1}{2}$ component $b_4$ 
is assumed to have a mass around 550~GeV\footnote{The cross-section for 
$t_4\bar{t_4}$ pairs is similar, since the mass-splitting between $b_4$ 
and $t_4$ cannot be much greater than about 50~GeV, but this 
cross-section has not been shown here since it is not relevant for our 
analysis.}. Finally, in the box marked ($c$) on the right, we have 
plotted cross-sections for the UED(5) model, where the particle spectrum 
lies in the vicinity of around 500~GeV. Although these cross-sections 
also vary quite dramatically with the choice of parameters, phase space 
considerations are rather similar for all the plots in 
Figure~\ref{fig:CrossSections}, and hence the apparent fact that for 
similar particle masses, the UED cross-sections tend to be larger than 
those in other models, happens to be true for most choices of 
parameters.

\bigskip\noindent The curves shown in Figure~\ref{fig:CrossSections} are 
{\it solely} intended to illustrate the general trend for heavy particle 
production through strong interactions as the centre-of-mass energy 
$\sqrt{s}$ of the LHC increases. The absolute values of the 
cross-sections in the three boxes vary widely depending on the parameter 
space. One should not, therefore, jump to conclusions such as imagining 
the SM(4) cross-sections to be generically smaller than the LH(T) 
cross-sections, which seems to be indicated by the central box ($b$) -- 
this is by no means true for other choices of parameters. However, it is 
reasonable to draw the conclusions that
\begin{itemize}
\item at $\sqrt{s} =10$~TeV, the overall cross-sections for strong 
production of new, heavy particles lies in the wide range of about 100 
fb to about 10 pb; and
\item the variation in cross-section as one proceeds from 7 to 14 TeV is 
about one order of magnitude.
\end{itemize}
The first conclusion can be used to reinforce the contention made in the 
previous section that these new particles will be produced in huge 
numbers at the LHC, even with revised luminosity estimates. The second 
can be used to get a (very rough) picture of how our results will change 
as the machine energy increases.

\begin{figure}[htb]
\centerline{ \epsfxsize= 6.0 in \epsfysize= 5.0 in \epsfbox{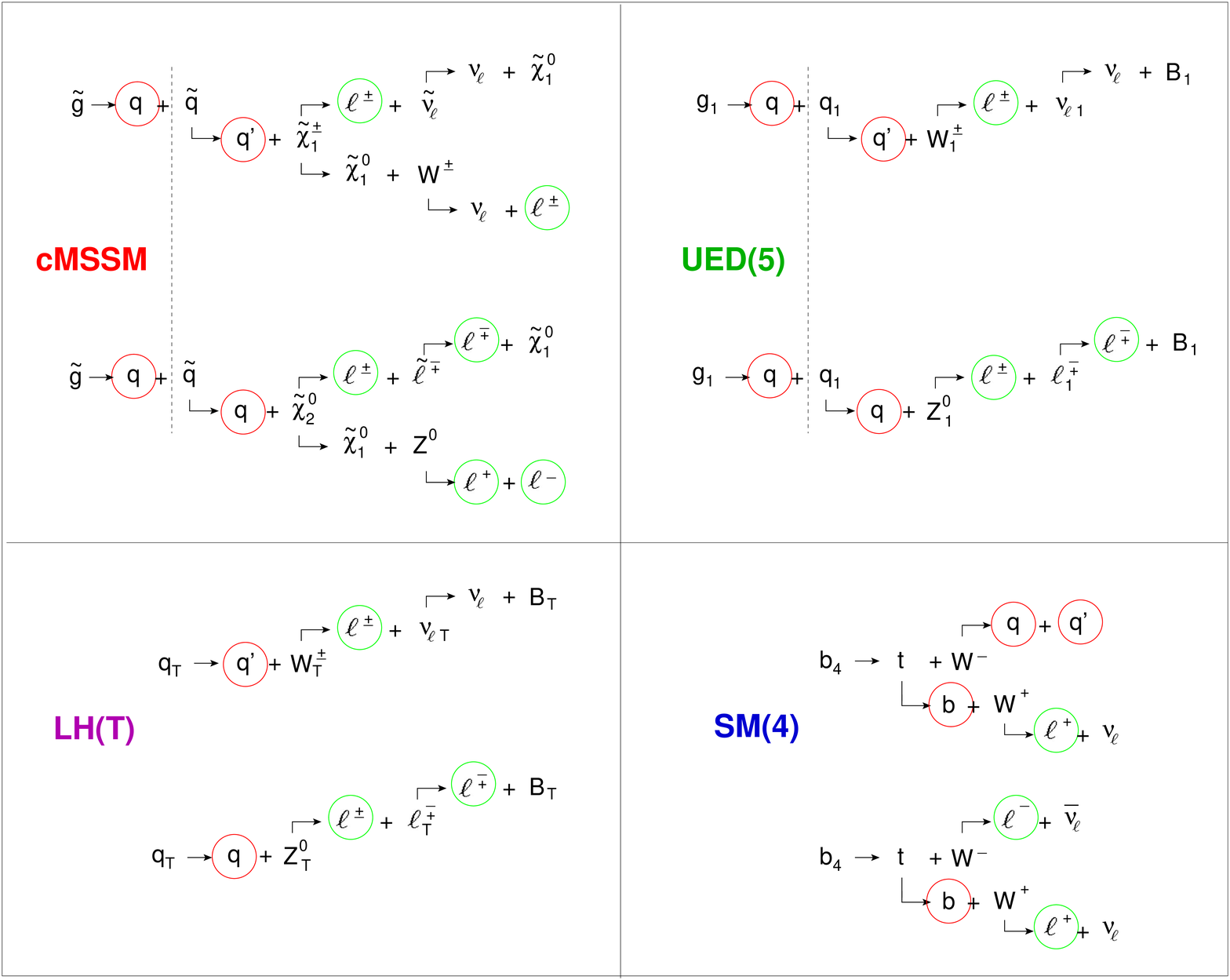} }
\caption{{\footnotesize\it Major cascade decay chains leading to 
multilepton signals with missing energy in $pp$ collisions at the LHC, 
in the four models under consideration. Final states are highlighted by 
encircling in red for jets and green for leptons. Other stable final 
states are invisible and lead to MET. The cascading process may also 
start from the state immediately to the right of the vertical (broken) 
line(s). In each box, the upper chain leads to {\em one} lepton, jet(s) 
and MET, while the lower chain leads to {\em two} leptons, jet(s) and 
MET.}}
\label{fig:Cascades}
\end{figure}

\bigskip\noindent Once such heavy new particles are produced in 
proton-proton collisions, they will undergo cascade decays as permitted 
by the conservation laws in the theory, leading, as we have already 
asserted, to multi-lepton signals. These multi-lepton signals 
accompanied by hadronic jets and large amounts of missing transverse 
energy (MET) are best understood with reference to the decay channels 
depicted in Figure~\ref{fig:Cascades}. These are now described in a 
model-wise manner.

\vskip -5pt

\noindent {\it Cascade decays in the} cMSSM : Depicted in the box marked 
`cMSSM', on the upper left side of Figure~\ref{fig:Cascades}, these 
cascade decays start from the gluino $\widetilde{g}$, though it is also 
possible to start on the right of the vertical broken line from the 
squark $\widetilde{q}$ if that is the produced particle. Be it a gluino 
or a squark, two kinds of cascade are depicted in the figure. The decay 
of the gluino to the squark(s) proceeds through the dominant 
$\widetilde{g} \to \widetilde{q} + q$ channel(s) with branching ratio 
unity, but the squark(s) may decay in various ways. If they decay to 
chargino $\widetilde{\chi}_1^\pm$ states, the cascade decay follows the 
upper chain depicted in the figure, culminating in a final state with 
{\it one} lepton, jets and MET from the $\nu_\ell$ and the LSP 
$\widetilde{\chi}_1^0$. The chargino $\widetilde{\chi}_1^\pm$ itself may 
decay through two major channels, depending on its composition in terms 
of gaugino and Higgsino states. For example, if the chargino 
$\widetilde{\chi}_1^\pm$ is purely a Wino $\widetilde{W}^\pm$ and the 
LSP $\widetilde{\chi}_1^0$ is purely a Bino $\widetilde{B}^0$, then the 
$SU(2)_L\otimes U(1)_Y$ electroweak symmetry precludes the existence of 
a $\widetilde{\chi}_1^\pm$--$\widetilde{\chi}_1^0$--$W^\pm$ vertex, and 
hence one of the cascade decays depicted in the figure will not occur. 
However, since the final state is the same ($\ell^\pm$ + MET) the 
chargino branching ratio to one lepton is effectively unity.

\bigskip\noindent It is also possible for the squark(s) to decay into 
one of the neutralino states $\widetilde{\chi}_i^0$ ($i = 1,2,3,4$). If 
it decays directly to the LSP\footnote{We note here that in the cMSSM, 
the lightest neutralino $\widetilde{\chi}_1^0$ is always the LSP in the 
parameter space allowed by experimental constraints.}, i.e. 
$\widetilde{q} \to q + \widetilde{\chi}_1^0$, then, although the channel 
will be kinematically favoured, no leptons will be generated. On the 
other hand it is also possible for the squarks to decay to a heavier 
neutralino states, such as the $\widetilde{\chi}_2^0$ depicted in the 
figure, or even the heavy states $\widetilde{\chi}_3^0$ or 
$\widetilde{\chi}_4^0$. These branching ratios will be very much 
dependent on the choice of parameters. However, cascade decays of a 
heavy neutralino will lead -- again through two possible decay chains as 
depicted -- to a final state with $\ell^+\ell^-$ and MET.

\bigskip\noindent The two major decay chains above result in final 
states with either $\ell^\pm$ + MET + jets, or $\ell^+\ell^-$ + MET + 
jets. There are other, more exotic, possibilities. For example, we can 
have a decay chain of the form
$$
\widetilde{q} \longrightarrow  \widetilde{\chi}_2^\pm + \dots 
              \longrightarrow  \widetilde{\chi}_3^0   + \dots  
              \longrightarrow  \widetilde{\chi}_1^\pm + \dots  
              \longrightarrow  \widetilde{\chi}_1^0   + \dots
$$
where one can get one lepton from each of the chargino $\leftrightarrow$ 
neutralino decays, i.e. three leptons in all. However, the nature of the 
mass spectrum in the cMSSM makes it difficult to have other 
possibilities, since the $\widetilde{\chi}_4^0$ and the 
$\widetilde{\chi}_2^\pm$ are practically degenerate, as are the 
$\widetilde{\chi}_2^0$ and the $\widetilde{\chi}_1^\pm$. Even the 
four-step decay chain shown here is kinematically disfavoured except in 
some small patches of the parameter space and may be considered a rare 
process even there. We have chosen, therefore, to concentrate mostly on 
the decay chains depicted in Figure~\ref{fig:Cascades}.

\begin{figure}[htb]
\centerline{\epsfxsize=4.3in\epsfysize=3.0in\epsfbox{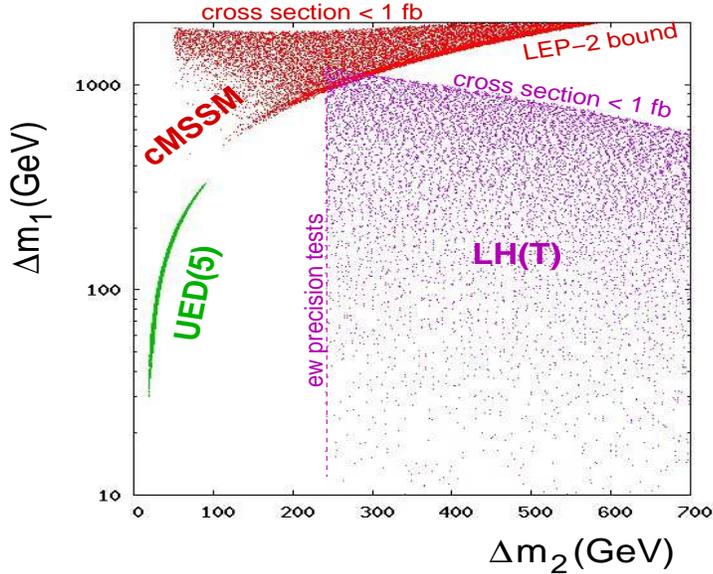}}
\caption{\footnotesize\it Illustrating mass difference patterns 
responsible for three of the four models under consideration. The 
variables plotted along the axes are as follows: }
\label{fig:Spectrum}
\end{figure}

\bigskip\noindent Once we decide to concentrate on the one-lepton chain 
$\ell^\pm$ + MET + jets and the two-lepton chain $\ell^+\ell^-$ + MET + 
jets, there are three possibilities after a production of a pair of the 
strongly-interacting particles $\widetilde{g}$ and/or $\widetilde{q}$, 
viz.,
\begin{itemize}
\item both decay through the one-lepton chain, so that the final state 
is ($2\ell$ + MET + jets);
\item one decays through the one-lepton chain while the other decays 
through the two-lepton chain, so that the final state is ($3\ell$ + MET 
+ jets);
\item both decay through the two-lepton chain, so that the final state 
is ($4\ell$ + MET + jets).
\end{itemize}
The first option, which has a substantial background from Drell-Yan 
production of leptons with some MET which can be due to various 
causes\footnote{e.g. radiation of a $Z$ which decays invisibly to 
neutrinos, or radiation of a number of soft gluons which form soft, 
undetected jets.} is not the focus of our attention in this work. It is 
the second and third options, i.e. the ($3\ell$ + MET + jets) and the 
($4\ell$ + MET + jets) signals which interest us, as their SM 
backgrounds are negligible. The jet multiplicity will depend on the 
initial state as well as the leptonic content, apart from possibilities 
for jet splitting and jet merging. However, we can broadly conclude that 
the energy of the jets will be controlled by the splitting between 
$M(\widetilde{g})$ and $M(\widetilde{q})$ and between $M(\widetilde{q})$ 
and $M(\widetilde{\chi}_1^\pm) \approx M(\widetilde{\chi}_2^0)$. To 
quantify this, we define a variable $\Delta m_1 = {\rm 
sup}\{M(\widetilde{g}),M(\widetilde{q})\} - 
M(\widetilde{\chi}_1^\pm)\}$. Similarly, the energy of the emergent 
lepton(s) will be controlled by the mass difference $\Delta m_2 = 
M(\widetilde{\chi}_1^\pm) - M(\widetilde{\chi}_1^0)$. A scatter plot of 
these mass differences is shown in Figure~\ref{fig:Spectrum}, where the 
red dots represent randomly chosen points in the cMSSM parameter space 
which are allowed by all laboratory constraints, including those from 
LEP-2, radiative $B$-decays, etc. The strong lower bound on $\Delta m_1$ 
is mostly controlled by the direct bound on the chargino mass from 
LEP-2, while the upper cut-off (near the top of the box) has been 
imposed by demanding that the cross-section for coloured sparticle 
production at the LHC should not fall below 1~fb at $\sqrt{s} = 10$~TeV. 
It is immediately obvious that the cMSSM spectrum in the allowed 
parameter space contains heavy gluino and squark states with masses 
around a TeV or more, while the gaugino states are relatively light. 
This accounts for the fact that $\Delta m_1$ is generically above 
500~GeV, while $\Delta m_2$ is generically below 600~GeV. It follows 
that jet energies in the range of many hundreds of GeV's are possible in 
the cMSSM, while the leptons will tend to be softer, being most likely 
around 200 -- 300 GeV. The very hard jets in this case will have ample 
energy to split into more jets and give a high jet multiplicity, and 
usually the leptons will all be well above any reasonable detection 
thresholds we may choose to apply, as a result of which they will be 
easily detected.

\bigskip\noindent {\it Cascade decays in the {\em UED(5)} model} : It is 
not for nothing that the UED(5) model has been dubbed `bosonic 
supersymmetry'\cite{UED5signals}. The $n = 1$ particle spectrum and 
decay chains in this model resemble supersymmetry very closely. Yet 
there are significant differences, as the depiction in 
Figure~\ref{fig:Cascades} shows. The upper right-hand box shows how a 
$g_1$ KK excitation can decay through a $q_1$ excitation into either a 
$W_1$ or a $Z_1$ KK excitation, whose further decays then yield one 
lepton or two leptons respectively, with the stable LKP, the $B_1$ 
(often denoted $\gamma_1$) escaping detection and contributing to MET. 
It should be noted at this point that in the UED(5) model, the $n = 1$ 
counterpart of the Weinberg angle is very small\cite{UED5spectrum} and 
hence one may say that the LKP $\gamma_1$ is practically all $B_1$, 
while the orthogonal state $Z_1$ is practically all $W_1^{(3)}$. This 
represents a rather different situation from the cMSSM, where the LSP 
$\widetilde{\chi}_0^1$ is almost always a mixture of the $\widetilde{B}$ 
and the $\widetilde{W}^{(3)}$, except in some very small patches of the 
parameter space. The dominantly $U(1)_Y$ nature of the LKP in the UED(5) 
model means that vertices like $W_1^\pm W^\mp \gamma_1 \approx W_1^\pm 
W^\mp B_1$ are absent, unlike their cMSSM counterparts. At the same 
time, there is no analogue of the cMSSM process $\widetilde{\chi}_2^0 
\to \widetilde{\chi}_1^0 + Z^0$ which is a supersymmetrization of the 
$ZZH$ vertex in the SM. This is because the LKP, the analogue of the 
$\widetilde{\chi}_1^0$, is almost wholly a $B_1$\cite{UED5spectrum}.

\vskip -10pt

\noindent {\footnotesize\it
\begin{itemize}
\item {\em cMSSM:} 
$\Delta m_1 = {\rm sup}\{M(\widetilde{g}) - M(\widetilde{\chi}_1^\pm), 
M(\widetilde{q}) - M(\widetilde{\chi}_1^\pm)\}$ and 
$\Delta m_2 = M(\widetilde{\chi}_1^\pm) - M(\widetilde{\chi}_1^0)$ 
\item {\em UED(5):} 
$\Delta m_1 = {\rm sup}\{M(g_1) - M(W_1^\pm), M(q_1) - M(W_1^\pm)\}$ and 
$\Delta m_2 = M(W_1^\pm) - M(B_1^0)$ 
\item {\em LH(T):} 
$\Delta m_1 = M(q_T) - M(W_T^\pm)$ and 
$\Delta m_2 = M(W_T^\pm) - M(B_T^0)$ 
\end{itemize}
\vskip -5pt
Here, $\Delta m_1$ represents an upper bound for jet energies and 
$\Delta m_2$ represents an upper bound for lepton energies.}

\bigskip\noindent Even lacking two of the important decay chains which 
are present in the cMSSM, it is still possible for the UED(5) to predict 
final states with ($3\ell$ + MET + jets) and ($4\ell$ + MET + jets) 
through the decay chains depicted in Figure~\ref{fig:Cascades}, i.e. 
through an intermediate lepton number-carrying $n = 1$ state, which is 
analogous to the slepton or the sneutrino state depicted in the cMSSM. 
However, we now note that there exists another important difference 
between the cMSSM and the UED(5) model. In the cMSSM, all fermion masses 
are equal (to $m_{1/2}$) at some high scale around $10^{16-17}$~GeV and, 
likewise, all boson masses are equal (to $m_0$) at the same high scale. 
Radiative corrections -- which depend crucially on the various couplings 
and conserved quantum numbers carried by these states -- ensure that 
these masses have very different running properties when calculated at 
the energy scale of the LHC, using renormalisation group (RG) equations. 
So widely do these vary that the mass spectrum at the LHC scale of 
around a TeV permits mass splittings as large as a TeV 
(Figure~\ref{fig:Spectrum} makes this quite explicit.). In the UED(5) 
model, however, effects of compactification of the higher dimension are 
observed at the LHC scale itself, where all KK excitations of level $n$ 
will have a common tree-level mass $nR^{-1}$, where $R$ is the radius of 
compactification. As in the cMSSM, this degenerate mass spectrum is 
split by radiative corrections, but this time since everything happens 
at the TeV scale, the radiative corrections remain small, as is only to 
be expected from perturbative effects. There is no RG running here to 
blow up the small mass-splittings into huge differences.

\bigskip\noindent What the above argument means in practice is that the 
overall mass difference between the heavy $g_1$ and the LKP $B_1$ is not 
very large. In fact, a good rule of thumb is to take this difference as 
around 15\% of the generic $n = 1$ mass $R^{-1}$. As particles which are 
much heavier than about 1.5~TeV would have small production 
cross-sections at the LHC, the mass splitting between $n = 1$ states 
cannot be much more than about 225~GeV. The mass splittings between 
$g_1$ and $q_1$, between $q_1$ and $W_1$ and between $W_1$ and $B_1 
\approx \gamma_1$ must be less than this. Since this mass splitting has 
to account for the transverse energy of the final state jets and 
leptons, as well as the MET, it is clear that none of these quantities 
can be expected to peak anywhere above 100 GeV, though the occasional 
statistical fluctuation is, of course, always possible. The 
near-degenerate nature of the UED(5) spectrum is nicely illustrated in 
Figure~\ref{fig:Spectrum}. Here, as before, we define a pair of 
mass-splittings $\Delta m_1 = {\rm sup}\{M(g_1) - M(W_1^\pm),M(q_1) - 
M(W_1^\pm)\}$ and $\Delta m_2 = M(W_1^\pm) - M(B_1)$ and show a scatter 
plot of randomly chosen points in the UED(5) parameter space. These are 
represented by the green dots in Figure~\ref{fig:Spectrum}. It is clear 
that these splitting are much smaller than the splittings in the cMSSM, 
as a result of which the green dots form a narrow sliver in the $\Delta 
m_1$--$\Delta m_2$ plane, which {\it does not overlap} the cMSSM region 
at all.

\bigskip\noindent Before passing on to the next model, it is important 
to discuss the KK modes for $n \geq 2$. Since the lower bound on 
$R^{-1}$ can be as low as some 300--400~GeV\cite{UED5bounds}, it is 
easily possible to produce the $n = 2$ KK states (which have even 
KK-parity) as resonances. The mass of such a resonance will be roughly 
$2R^{-1}$, which is the same as that of a pair of $n = 1$ KK states 
having masses $R^{-1}$ each. On the other hand $n =3$ resonances, which 
have odd KK-parity, have to be pair-produced, which requires a 
centre-of-mass energy of around $6R^{-1}$, which puts it in the vicinity 
of 2~TeV or higher. The sharp falling-off of PDFs at such high energies 
(especially if LHC is unable to reach the design goal of 14~TeV) ensures 
that cross-sections for pair-production of $n = 3$ states will be too 
small to merit further discussion. However, $n =2$ resonances have been 
discussed in the literature\cite{UED5n2} as a direct test of the UED(5) 
model vis-\'a-vis supersymmetry \footnote{which, being $N = 1$ 
supersymmetry in order to have chiral fermions, has no analogue of the 
higher $n$ KK excitations.}. In this work, however, we find that it is 
possible to distinguish the UED(5) signals from the cMSSM ones 
considering the $n = 1$ states alone, and hence we do not pursue the 
issue of higher $n$ states any further. Such signals, if found, may be 
taken as complementary to the results and arguments presented in this 
article.

\bigskip\noindent {\it Cascade decays in the {\em LH(T)} model} : Little 
Higgs models\cite{LittleHiggs} have found their own niche in the lore of 
particle physics as they provide a solution to the hierarchy problem 
which is intermediate between that provided by supersymmetry and brane 
world models with either large, flat or small, warped extra dimensions. 
The introduction of two complementary gauge groups $SU(2)_1\times 
U(1)_1$ and $SU(2)_2\times U(1)_2$ which are diagonally broken to the SM 
gauge group $SU(2)_L\times U(1)_Y$ enables us to cleverly assign quantum 
numbers to bosons and fermions transforming under these in such a way as 
to cancel the leading quadratic contributions to the Higgs boson 
self-energy\footnote{However, this cancellation occurs only to first 
order, so that the hierarchy problem reappears at higher orders. The 
model is thus able to postpone the hierarchy problem to a higher scale 
of around 10~TeV, which is beyond the accessible energy of the LHC, but 
not to remove it altogether.}. To keep the extra fermions from making 
unacceptably large contributions to electroweak corrections to the $W$ 
and $Z$ boson masses, direct couplings have to be forbidden by the 
introduction\cite{LHTmodel} of a conserved quantum number called 
$T$-parity. It turns out that in these LH(T) models, all SM particles 
$P$ participating in electroweak interactions have $T = +1$ and possess 
a $T=-1$ partner $P_T$. Note that there is no $T$-odd partner for 
gluons, since they do not participate in electroweak interactions. The 
$T$-odd partners are generally quite heavy, satisfying\cite{LHTspectrum} 
\begin{equation} M(W_T) = gf \qquad\qquad M(B_T) = \frac{g'f}{\sqrt{2}} 
\qquad\qquad \left[\mathbb{M} (q)_T\right]_{ij} = \kappa_{ij}f 
\label{eqn:LHTmasses} \end{equation} where $f$ is a common mass scale 
(around a TeV) and the $\kappa_{ij}$ form a matrix of Yukawa couplings. 
Keeping the $T$-odd quarks $q_T$ (and, separately, $T$-odd leptons 
$\ell_T$) degenerate helps avoid large FCNC effects through a GIM 
mechanism. The mass splitting between the $W_T$ and the $B_T$ works out 
to about $2.75 \ M(B_T)$. Given that the electroweak precision 
bounds\cite{LHTspectrum} on LH(T) models lead to $M(B_T) > 80$~GeV, it 
follows that $M(W_T) - M(B_T) > 220$~GeV, which is quite substantial.

\bigskip\noindent It is apparent from Eqn.~(\ref{eqn:LHTmasses}) that 
the mass spectrum in the LH(T) model is not so tightly determined as is 
the case in the cMSSM and in the UED(5) models. The base parameters in 
the theory are the scale $f$, the common $q_T$ mass $M(q_T)$ and the 
common $\ell_T$ mass $M(\ell_T)$, which can be varied at will. Not every 
mass scenario in the model, therefore, leads to $3\ell$ and $4\ell$ 
signals. The ones which do so are compatible with the sketch shown in 
Figure~\ref{fig:Cascades} (box on the second row, to the left, marked 
LH(T)) and have a hierarchy
\begin{equation}
M(q_T) > M(W_T), M(Z_T) > M(\ell_T) > M(B_T)
\label{eqn:Hierarchy}
\end{equation}
which leads to cascade decays with multi-lepton states. This is not to 
say that the other possibilities (e.g. having the $\ell_T$ heavier than 
the $W_T$ or $Z_T$) are considered in any way less favoured. It is 
simply that such scenarios will not lead to multi-lepton scenarios, and 
hence are not relevant for the inverse problem in the limited sense 
under consideration. For the mass pattern which does give multi-lepton 
signals, however, Figure~\ref{fig:Cascades} shows that the decay pattern 
for the LH(T) model is very similar to the pattern depicted for the 
UED(5) model to the right of the vertical (broken) line.

\bigskip\noindent On the whole, the LH(T) scenario may be expected to 
lead to somewhat smaller cross-sections than the cMSSM or the UED(5) 
because of the absence of any counterpart of the $\widetilde{g}$ or the 
$g_1$, and hence, no counterpart of the $\widetilde{g} \widetilde{g}$ 
and $\widetilde{q}\widetilde{g}$ (or $g_1g_1$ and $q_1g_1$) processes. 
This, by itself, cannot be used as a distinguishing feature, of course, 
since the LH(T) cross-section for a smaller mass range is comparable to 
the cMSSM or UED(5) cross-section for a larger mass range. However, it 
can be used in conjunction with other parameters, as will be seen in the 
final section. Of more immediate interest is the fact that the mass 
splittings $\Delta m_1 = M(q_T) - M(W_T)$ and $\Delta m_2 = M(W_T) - 
M(B_T)$, which govern the energy of emergent jets and leptons 
respectively, are neither constrained to be as small as in the UED(5) 
case, not required to be as large as in the cMSSM. These intermediate 
values are plotted as a scatter plot in Figure~\ref{fig:Spectrum}, where 
it is apparent that, except for a small region of overlap with the cMSSM 
region, the LH(T) model largely occupies a different region in the 
$\Delta m_1$---$\Delta m_2$ plane. This difference can be effectively 
exploited in building discriminators, as we show in the next section.

\bigskip\noindent {\it Cascade decays in the {\em SM(4)} }: As mentioned 
in the Introduction a sequential fourth generation in the SM is still a 
possibility if there is a modest mass-splitting in the quark sector 
which is matched by a corresponding mass-splitting in the lepton sector, 
in accordance with Equation~(\ref{eqn:Sparameter_in_SM4}). In fact, 
balancing the constraints from the $S$ and the $T$ parameters, which 
tend to work in opposite directions, one concludes that the mass 
splitting between $t_4$ and the $b_4$ can be anything in the range from 
0--55 GeV at 95 \% CL. Constraints coming from current data on 
$B^0$-$\bar{B}^0$ mixing and $Z\to b\bar{b}$ are not incompatible with 
this level of mass-splitting\cite{SM4lowenergy}.

\bigskip\noindent In the presence of a fourth generation the $3 \times 
3$ Cabibbo-Kobayashi-Maskawa (CKM) matrix $V$ requires to be extended to 
a $4 \times 4$ matrix, where the mixing elements with the first two 
generations are constrained by unitarity and other considerations to be 
practically zero. There can be more substantial mixing between the third 
and fourth generations\footnote{Which is consistent with the hierarchy 
of mixing observed in the first three generations, viz. neighbouring 
generations mix more.} with the only constraint on $V_{34}$ coming from 
from the unitarity of the $4\times 4$ mixing matrix. This forces 
$V_{44}$ to be close to 1, and $V_{34} \leq 0.15$ at 95 \% CL. For 
values which saturate the latter bound, both the $t_4$ and $b_4$ will 
dominantly decay through this mixing to two-body final states through 
$t_4 \to b + W^+$ and $b_4 \to t + W^-$, since the CKM-favoured decays 
$t_4 \to b_4 + W^+$ or $b_4 \to t_4 + W^-$ are kinematically impossible 
when $|M(t_4) - M(b_4)| < 55~{\rm GeV} < M_W$, and can only be realised 
in three-body decays with a heavy $W$-propagator. Of course, the precise 
values of $|V_{34}| \approx |V_{43}|$ is not important for collider 
studies. They will only affect the lifetimes of the fourth-generation 
quarks and not their decay patterns, since the two-body decay will 
dominate over the three-body one unless this element $|V_{34}| \approx 
|V_{43}|$ assumes an unnaturally small value of $10^{-4}$ or less.

\bigskip\noindent Once we realise that the $t_4$ and $b_4$ quarks will 
decay into ordinary $b$ and $t$ quarks, it is clear that direct searches 
at Tevatron can put stringent limit on their masses. In fact, such 
limits do exist\cite{SM4bounds}, and are about $M(t_4) > 325$~GeV and 
about $M(b_4) > 355$~GeV, assuming that they decay dominantly to the 
third generation quarks. There is no upper limit on the masses of the 
$t_4$ and $b_4$, except for the quantum field theoretical constraint 
that if they are more massive than about 600~GeV, the corresponding 
Yukawa couplings would be large and would then quickly hit the Landau 
pole\cite{SM4facts}.

\bigskip\noindent The presence of a fourth generation of quarks with 
masses in the range 350--600~GeV could easily act as a spoiler for new 
physics signals in the trilepton and four-lepton channels. The box on 
the lower right side of Figure~\ref{fig:Cascades} shows the decay chains 
which are responsible for this. As in the other boxes, the upper chain 
gives rise to a $\ell$ + MET + jets signal, while the lower chain gives 
rise to a $\ell^+\ell^-$ + MET + jets signal. We notice that both these 
chains arise from the decay of a $b_4$ quark. If we produce a $t_4$ 
quark, it will always decay to a $b$-quark and a $W$-boson, producing 
not more than a single lepton when the $W$-boson decays\footnote{Of 
course, a small number of isolated leptons will also arise when the 
$b$-quark decays semi-leptonically, but this can be treated as a 
sub-leading effect.}. Thus, the decay of a $t_4 \bar{t}_4$-pair will 
lead at most to a $2\ell$ + MET + jets signal, which is not under 
consideration in this work. Thus, the $t_4$ has no role to play in the 
subsequent discussion and will not be considered any further.

\bigskip\noindent Being a minimal extension of the SM, the SM(4) has no 
counterpart of the LSP ($\widetilde{\chi}_1^0$), the LKP ($B_1$) or LTP 
($B_T$). Hence the only mass-splitting that is relevant is $M(b_4) - 
m_t$, which is simply linear in $M(b_4)$, varying from about 
180--425~GeV. However, this mass difference does not represent the 
energy of any jet or lepton {\it per se}, but is shared among five light 
particles according to the kinematic configuration. Accordingly, we do 
not show any scatter plot in Figure~\ref{fig:Spectrum} corresponding to 
the SM(4). It turns out, however, that the very lack of a characteristic 
spectrum for the SM(4) makes it difficult -- though not impossible -- to 
disentangle from the other models. This will be discussed in the 
following sections.

\bigskip\noindent Though we concentrate on the four models described 
above, trilepton and four-lepton signals are also possible in some other 
models. The obvious ones are extensions or variations of the above 
models, such as the many possibilities with 
supersymmetry\cite{SUSYreview} --- non-universal masses, focus-point 
evolution of masses, even the unconstrained MSSM, as well as the 
possibility of $R$-parity violation through a $LL\bar{E}$ coupling. 
Other possibilities include the extension of the UED(5) model to 6 
dimensions\cite{UED6}, or a composite model with appropriate couplings. 
In fact, it is always possible to build up a model with suitably chosen 
fields and interactions which would produce the given signals. However, 
such ad hoc creations would lack the motivational advantages of the four 
scenarios discussed in this work. Hence, we have not made any attempt to 
study the ($3\ell$ + MET + jets) and ($4\ell$ + MET + jets) signals in 
any models other than the four enlisted above.

\bigskip\noindent It is necessary to add a caveat to the previous 
paragraph. There is always a possibility that when the LHC data become 
available, we shall see signals in the ($3\ell$ + MET + jets) and 
($4\ell$ + MET + jets) channels which do not match in kinematic and 
other profiles with any of the four models considered in this article. 
In that case, it is precisely the kind of `exotic' possibilities 
mentioned in the last paragraph, to which we would require to turn for 
an explanation. Our purpose is not, therefore, to belittle these 
alternative models or to be dismissive of their relevance, but simply to 
note that the present study -- a first of its kind -- is necessarily 
limited in scope, and does not take all these models into account.

\section{Model Discriminators}

\noindent At this point we expect that the reader will have been 
convinced that ($a$) different models, such as the cMSSM, UED(5), LH(T) 
and SM(4), and perhaps a few others as well, will indeed, contribute to 
new physics signals in the ($3\ell$ + MET + jets) and the ($4\ell$ + MET 
+ jets) channels, and that ($b$) the kinematic footprint of different 
models are likely to be different, given the different mass patterns 
shown in Figure~\ref{fig:Spectrum}. However, in order to be precise, we 
must develop ($i$) numerical measures of these footprints, and ($ii$) a 
systematic way of studying them, which would lead to economic and 
efficient ways of discriminating between different models of new 
physics. These are the goals of the present discussion.

\bigskip\noindent At this point, it is a good idea to quickly take stock 
of the main tools one can use for kinematic analysis. These are really 
of three kinds, viz.
\begin{itemize}
\item {\it kinematic distributions}, i.e. differential cross-sections in 
kinematic variables;
\item {\it cross-sections after putting kinematic cuts}, which are the 
same thing in another avatar;
\item {\it multiplicity count}, i.e. counting the number of jets, 
leptons, etc. which emerge after putting appropriate kinematic cuts.
\end{itemize}
The maximum amount of information will, of course, lie in the first 
option, but this is inconvenient when scanning over the parameter space 
of an underlying theory. Of course, once there exists a kinematic 
distribution from the experimental data, one can treat that as a 
standard and check theoretical predictions against it -- using some 
fitting procedure such a maximum likelihood fit or a Bayesian analysis. 
Again, if there is a substantial SM cross-section, we can perhaps treat 
that as a standard and check theoretical predictions against it. For the 
trilepton and four-lepton signals, however, we have neither, and hence, 
there is no single numerical index to tell us whether a given 
theoretical distribution is good or bad. It is more convenient, 
therefore, to use the second and third options, in which every point in 
the parameter space maps into a single number (the cross-section) or a 
few numbers (the multiplicities). If we plot two such numbers along the 
axes of a `signature space', the variation of parameters will generate a 
region in the `signature space'. Overlapping regions would then 
correspond to non-distinguishability of models; non-overlapping regions 
will mean that the models are distinguishable.

\begin{figure}[htb]
\centerline{ \epsfxsize= 6.5 in \epsfysize= 5.0 in \epsfbox{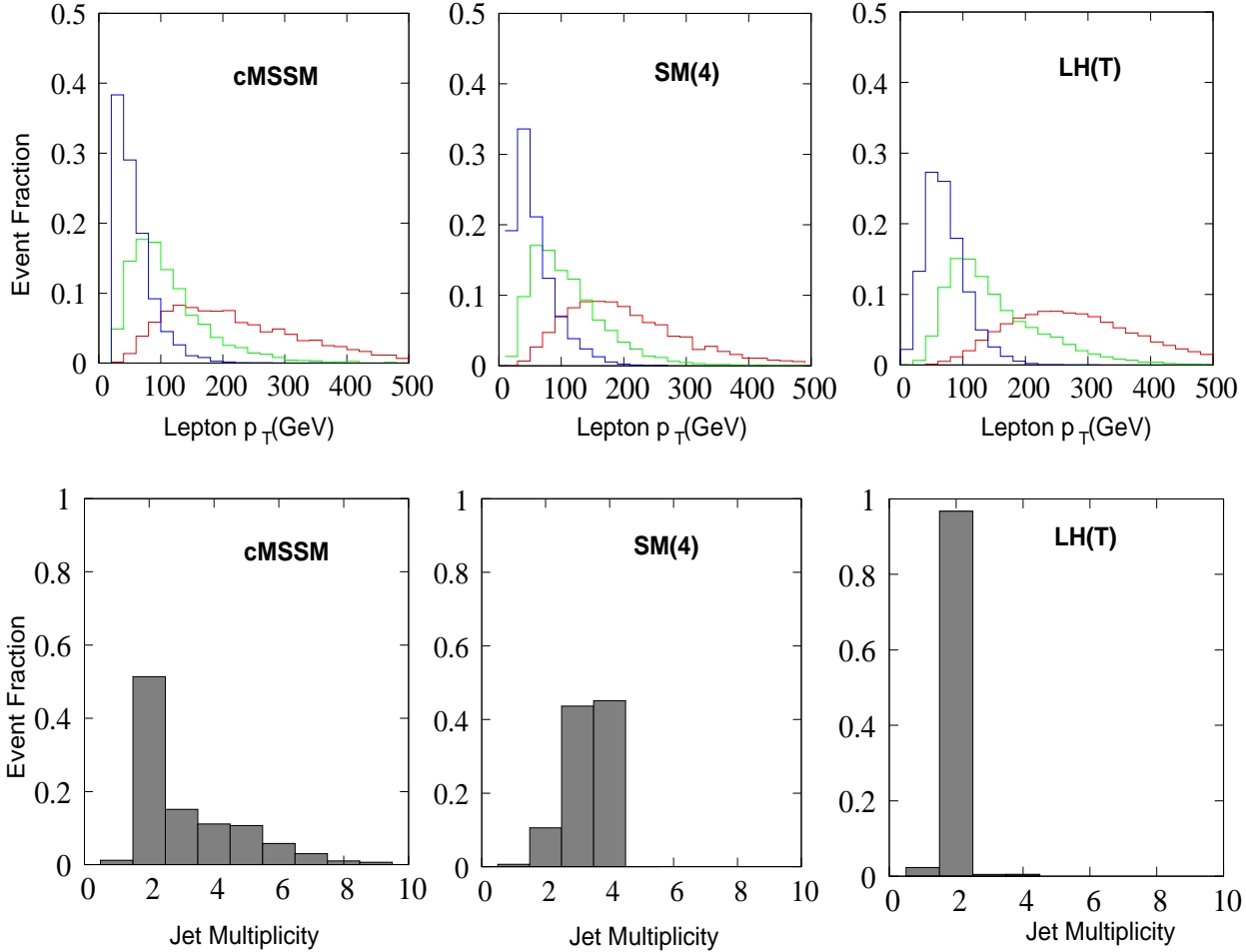} }
\caption{{\footnotesize\it Sample kinematic distributions for one point 
in the parameter space (see text) of each of the three models {\em 
cMSSM, SM(4)} and {\em LH(T)}. The three upper boxes show the $p_T$ 
spectrum for the leptons in a trilepton signal, with the red, green and 
blue histograms corresponding to the three leptons in order of hardness. 
The three lower boxes, show, for the same choices of parameters, the jet 
multiplicity in each of these models..}}
\label{fig:Distributions}
\end{figure}

\bigskip\noindent To make the general discussion above more concrete, in 
Figure~\ref{fig:Distributions}, we have plotted some kinematic 
distributions (normalized to unity) in the upper half and some 
multiplicity counts (normalized to unity) in the lower half, for a 
($3\ell$ + MET + jets) signal at the LHC, running at 10~TeV. The upper 
boxes show the transverse momentum $p_T$ of the three leptons (ordered 
according to their $p_T$) in the case of the cMSSM, SM(4) and the LH(T) 
models respectively, reading from left to right. Parameters in each of 
the models have been chosen arbitrarily in order to show that the lepton 
$p_T$ distribution can look quite similar in all the models. There are 
minor differences between the graphs shown in 
Figure~\ref{fig:Distributions}, but after detector smearing effects are 
taken into consideration, these are sure to be quite indistinguishable. 
A mere study of the $p_T$ spectrum of the leptons will not, therefore be 
able to discriminate between these three models.

\bigskip\noindent For transparency, in Figure~\ref{fig:Distributions}, 
the parameters for each model are:
$$
\begin{array}{rcl}  
{\rm cMSSM} & : & m_0 = 150~{\rm GeV}, \ m_{1/2} = 650~{\rm GeV}, A_0 = 0, 
\tan\beta = 10, \mu > 0 \\ \\
{\rm SM(4)}  & : & M(b_4) = 600~{\rm GeV} \\ \\
{\rm LH(T)}  & : & M(W_T) = 650~{\rm GeV}, M(B_T) = 150~{\rm GeV}, M(\ell_T) = 560~{\rm GeV}, M(q_T) = 1~{\rm TeV}
\end{array}
$$
However, there are many points in parameter space where similar 
distributions between models may be obtained, so this particular choice 
of parameters has no more significance than the fact that it serves to 
illustrate the argument. We have not shown the UED(5) case in 
Figure~\ref{fig:Distributions}, because there the leptons will tend to 
be much softer than the ones shown in the figure. However, that does not 
mean that the UED(5) model is distinguishable by these alone, since soft 
leptons can arise in some of the other models as well, for appropriate 
parameter choices.

\vskip -5pt

\noindent The three boxes in the lower half of 
Figure~\ref{fig:Distributions} show the jet multiplicity in each of the 
three models as indicated, for the {\it same} choice of parameters as in 
the upper boxes. For the record, these jets have been generated using 
PYTHIA, with appropriate hadronization and fragmentation as implicit in 
that software, and then collected into jets using the inbuilt toy 
calorimeter subroutine PYCELL. Though this is not a very sophisticated 
jet-making tool, it is known to be tolerably accurate, and in any case, 
a glance at the figure will show that smearing or no smearing, these 
multiplicity distributions are vastly different in all the three cases. 
Physically, this is not so difficult to understand:
\begin{itemize}
\item For the cMSSM, the ($3\ell$ + MET + jets) signal arises when there 
is a chargino-mediated decay on one side and a neutralino mediated decay 
on the other side. The dominance of the $\widetilde{q} \widetilde{q}^*$ 
production mode is reflected in the fact that the maximum number of 
events have 2 jets. The few single jet events can be variously 
attributed to loss of a jet through merging, extreme softness, passage 
down the beam pipe or excessive scatter of hadronic clusters. The 
next-most dominant process is $\widetilde{q} \widetilde{g}$ production, 
which leads to 3 jets, and this is indeed the next highest bar in the 
histogram. Finally $\widetilde{g}\widetilde{g}$ production, which is 
sub-dominant, provides most of the 4 jet events. Higher multiplicities 
arise when the $W^\pm$ bosons arising from 
$\widetilde{\chi}_1^\pm/\widetilde{\chi}_2^0$ cascades decay 
hadronically.
\item For the SM(4), three-leptons are found when three of the $W$ 
bosons in the chain decay leptonically, and one decays hadronically, 
i.e. mostly into two jets. Taken together with the two $b$-jets produced 
at the first step in the cascade, one would expect all trilepton events 
to be accompanied by 4 jets. Indeed the largest number of events do have 
4 jets, but many of these merge or are otherwise lost to give 
substantial numbers of events with 2 and 3 jets. Single jet events are 
comparatively rare, as is only to be expected, since that would involve 
loss or merging of 3 of the 4 jets.
\item For the LH(T), on the other hand, there are only two hard jets 
which arise from the decay of the parent $q_T$ (or $\bar{q}_T$). There is 
a little bit of merging, etc., but hardly any fragmentation effects to 
speak of.
\end{itemize} 
\noindent These jet multiplicities immediately indicate to us how to go 
about defining a discriminator variable. Let us consider the jet multiplicity
distributions of Figure~\ref{fig:Distributions} and define
\begin{equation}
D = \frac{\rm number \ of \ trilepton \ events \ with \ 3 \ or \ more \ jets}
         {\rm number \ of \ trilepton \ events \ with \ 2 \ jets \ or \ less}
\label{eqn:Dparameter}
\end{equation}
A quick glance at the three boxes in the lower half of 
Figure~\ref{fig:Distributions} shows that $D$ will be around unity for 
the cMSSM, much larger than unity for the SM(4) and very small for the 
LH(T). The fractional values given in the figure can be converted into 
numbers by multiplying by the total number of events (which is going to 
cancel in the ratio anyway). The exact figures are 0.9 for the cMSSM, 
7.9 for the SM(4) and 0.01 for the LH(T), and they span almost three 
orders of magnitude. Unfortunately, this wide separation between values 
of $D$ calculated in the three models is not always the case as we scan 
over the parameter space. For example, in LH(T) models, as the mass of 
the $q_T$ is brought down from 1~TeV to close to the $W_T$ mass, the 
jets produced in $q_T \to W_T$ decay will tend to get softer and will 
fragment more. In this case, the value of $D$ calculated in the LH(T) 
model will increase substantially. It may not overlap with the SM(4), 
but it can be expected to overlap with cMSSM events in some part of the 
cMSSM parameter space where the $\widetilde{q} \widetilde{q}^*$ 
dominates the others.

\bigskip\noindent Nevertheless, $D$ is a good and robust discriminator, 
because of several reasons, viz.
\begin{enumerate}
\item Since it essentially a {\it ratio} of cross-sections its value is 
independent of many multiplicative factors, such as luminosity, overall 
coupling constants, etc.
\item Since it involves two parts of the {\it same} cross-section, the 
major part of the PDF dependence also cancels. However, since the 
identification of jets and the jet formation algorithm depends on the 
kinemetics, the jet multiplicity may depend marginally on the colliding 
parton momentum fractions $x_1$ and $x_2$. This may in turn, induce a 
small, sub-leading, PDF dependence.
\item For the same reasons as above, the dominant parts of the QCD and 
electroweak corrections as well as factors due to multi-particle 
interactions, initial and final state radiation and detector 
efficiencies also cancel out.
\end{enumerate}
The robustness of $D$ is, however, affected by the requirement that in 
calculating the experimental value of $D$, the denominator should not be 
zero or even very small. To have a reasonable number in the denominator, 
the overall cross-section and luminosity again come into play, and 
through these, all the uncertain factors mentioned above. In this work, 
we have chosen to calculate $D$ and all $D$-like ratios only if both 
numerator and denominator are $\geq 5$, for a given luminosity 
benchmark. Thus, the parameter space accessible through $D$ and $D$-like 
variables gradually increases as the luminosity increases. This will 
become apparent in the next section, where we discuss our results.

\bigskip\noindent We must also take note of the fact that the 
information which enables us to discriminate between models is not 
entirely enshrined in the form of cross-section ratios like $D$ above. 
We can also use, with great profit, the actual magnitudes of the 
cross-section. For example, the production of a $g_1g_1$ pair yields a 
much greater cross-section than a $\widetilde{g}\widetilde{g}$ pair for 
the same mass range. As the coupling is just a strong coupling, this is 
a consequence of the different spin of the $g_1$ as opposed to the 
$\widetilde{g}$. Again, if we consider the LH(T) model, in general, 
there is no counterpart to the $\widetilde{g}\widetilde{g}$ and 
$\widetilde{g}\widetilde{q}$ (or $g_1g_1$ and $g_1q_1$) processes, so 
one would expect a smaller cross-section for the same mass range of the 
heavy $q_T$. Of course, the small cross-section in the LH(T) model for 
light $q_T$ production can be mimicked by small cross-sections in the 
cMSSM or UED(5) for heavy particle production. However, in the latter 
case, one expects to see larger mass splittings in the spectrum, which 
will show up in an appropriately-constructed ratio. Thus, combinations 
of a cross-section and a ratio may also lead to useful discriminators. 
However, when using a cross-section value as a discriminator, one has to 
realise that this is subject to variation due to all the effects 
mentioned above as cancelling for the $D$ parameter. Thus, only wide 
separations in cross-section should be taken seriously for 
discrimination purposes.

\bigskip\noindent Taking all the above considerations into account, and 
reflecting on the nature of the signals in the models under 
consideration, we have constructed a set of eight variables of 
event-counting nature, which, we believe, would retain the information 
required to discriminate between models at the LHC. These variable are 
listed in Table~1 below. Instead of using cross-sections, we have used 
event numbers assuming a given integrated luminosity ${\cal L}$ (which 
has been taken at different benchmark values in our numerical analysis). 
To define these event counting-type variables, we have used specific 
hardness criteria for both leptons and jets. These are
\begin{itemize}
\item A lepton will be considered {\it hard} if its transverse momentum 
satisfies $p_T^\ell \geq 50$~GeV ;
\item A jet will be considered {\it hard} if its transverse momentum 
satisfies $p_T^{\rm jet} \geq 150$~GeV.
\end{itemize}
The rationale for these hardness criteria may be obtained by glancing at 
Figure~\ref{fig:Spectrum} and noting that these would exclude most of 
the leptons and jets formed in the UED(5) model, for a large part of the 
parameter space. The other models depicted in the figure are not 
affected so strongly.

\begin{center} 
\begin{tabular}{cl} 
\hline 
Variable~~~ & ~~~Definition \\ \hline\hline 
$N_2^{(n)}$ & number of events with $n$ hard leptons and $\leq 2$ 
identifiable jets \\ 
$N_3^{(n)}$ & number of events with $n$ hard leptons and $\geq 3$ 
identifiable jets \\ 
$\widetilde{N}_0^{(n)}$ & number of events with $n$ leptons and no hard 
jet \\ 
$\widetilde{N}_1^{(n)}$ & number of events with $n$ leptons and $\geq 1$ 
hard jet \\ 
$\widetilde{\nu}$ & number of events with $\geq 1$ hard leptons and no 
hard jets \\ 
$\widetilde{\nu}_0$ & number of events with no hard lepton and no hard 
jets \\ 
$\widetilde{\nu}^\prime$ & number of events with $\geq 1$ hard lepton 
and $\geq 1$ hard jet \\ 
$\widetilde{\nu}^\prime_0$ & number of events with $\geq 1$ lepton and 
$\geq 1$ hard jet \\ \hline 
\end{tabular}

Table 1: {\footnotesize\it Event counting-type discriminators for 
$3\ell$ and $4\ell$ signals at the LHC, i.e. $n = 3,4$.}
\end{center}

\noindent In terms of the event-counting discriminators listed in 
Table~1, we are now in a position to define six ratio-type 
discriminators. These are
\begin{equation}
D_n = \frac{N_3^{(n)}}{N_2^{(n)}} 
\qquad\qquad 
\widetilde{D}_n = \frac{\widetilde{N}_1^{(n)}}{\widetilde{N}_0^{(n)}}
\qquad\qquad
\Delta_\ell = \frac{\widetilde{\nu}}{\widetilde{\nu}_0}
\qquad\qquad
\Delta_\ell^\prime = \frac{\widetilde{\nu}^\prime}{\widetilde{\nu}^\prime_0}
\label{eqn:Discriminators}
\end{equation}
for $n = 3,4$. All of these share the robustness properties defined for 
`$D$' above, but perform different functions so far as distinguishing 
between models is concerned. This is now discussed below, where we set 
out our naive expectations for each of the models under consideration. 
The reason these are called `naive' is because we do not know \'a priori 
how many extra jets will be formed by fragmentation, how many leptons 
will excite the trigger even though they come from semi-leptonic decays 
of $c$ and $b$ quarks in the jets, and such things. It is true that 
these effects are sub-leading in nature, but even so, one cannot 
completely neglect them, as the discrepancy between our naive 
expectations and the actual results obtained on running Monte Carlo 
simulations will presently show.

\bigskip\noindent As an illustration of the discriminating power of 
these ratio-type variables, let us consider the variable $D_3$, which is 
defined in Equation~(\ref{eqn:Discriminators}) as $D_3 = 
N_3^{(3)}/N_2^{(3)}$. This is equivalent to taking the jet multiplicity 
distribution corresponding to a trilepton signal, as shown in 
Figure~\ref{fig:Distributions} and partitioning it into two parts -- one 
with 0 -- 2 jets, and one with 3 or more jets. The ratio $D_3$ of these 
partitioned cross-sections is, in fact, identical to the variable `$D$' 
defined in Equation~(\ref{eqn:Dparameter}). Let us now see what our 
model-wise expectations for this variable $D_3 \equiv D$ are.
\begin{enumerate}
\item In the cMSSM (cf. Figure~\ref{fig:Cascades}), the number of parent 
partons available to form final state jets (apart from radiated gluons) 
can be 2, 3 or 4 depending on whether the initial state is 
$\widetilde{q}\widetilde{q}$, $\widetilde{q}\widetilde{g}$ or 
$\widetilde{g}\widetilde{g}$ respectively, with the largest 
cross-sections corresponding to 2 nascent jets. There will be some 
migration of events in the direction of increasing jet multiplicity as a 
result of jet-fragmentation and some migration in the direction of 
decreasing jet multiplicity because of jet-merging. One cannot tell \'a 
priori if one will dominate the other without doing an actual 
simulation. Nevertheless, it is reasonable to guess that the ratio of 
these cross-sections (or events numbers) will be of the order of unity, 
as indeed, it is for the parameter choice of 
Figure~\ref{fig:Distributions}.
\item In the UED(5) model, the argument is exactly identical with 
respect to parent partons, as the processes in question mimic the cMSSM 
processes almost exactly (cf. Figure~\ref{fig:Cascades}). Nevertheless, 
in this case, the near-degeneracy of the UED(5) spectrum causes the jets 
to be much softer than their cMSSM counterparts. In this case, the jet 
multiplicity will tend to be lower, partly because many of the jets will 
not be identifiable as such because of their extreme softness, and 
partly because soft jets tend to spread out more and merge more. We can, 
therefore, guess that $D_3$ will be smaller for the UED(5) model than it 
is for the cMSSM. To make a more precise estimate requires a detailed 
simulation.
\item In the SM(4), we have earlier seen that the parton-level count for 
nascent jets is 4, and all lower jet counts to arise from merging or 
jet-loss effects. Hence we expect $D_3$ to be larger than unity. Again, 
the exact figure will depend on the mass splittings and the jet 
identification and merging algorithms.
\item In the LH(T) model, again, we have earlier argued that the 
parton-level jet count is just 2, and hence larger number of jets will 
arise only as a result of fragmentation etc. Thus, we expect $D_3$ to be 
very small for the LH(T).
\end{enumerate}
Of course, all these estimates must be qualified by the scatter induced 
by parameter variation. Nevertheless, as the estimates are based on 
event toplogy rather than details of the spectrum, we expect a loose 
hierarchy in $D_3$ of the form $D_3^{\rm SM(4)} > D_3^{\rm cMSSM} > 
D_3^{\rm UED(5)} > D_3^{\rm LH(T)}$. Unless there is some major surprise 
in the trilepton signal, the experimental number is likely to eventually 
fall into one of the ranges covered by these different models.

\bigskip\noindent The behaviour and discriminating power of the $D_4$ 
variable is rather similar, and the only difference lies in the fact 
that it corresponds to a 4 lepton trigger. This makes it truly different 
only for the SM(4) case, where now all four $W$ bosons must decay 
through leptonic channels, and hence the parton-level expectation 
becomes 2 jets instead of 4. Thus, we expect $D_4$ to be small for the 
SM(4) unlike the $D_3$, which is large. Correlating a small $D_4$ with a 
large $D_3$ is, therefore, a characteristic of the SM(4).

\bigskip\noindent We can make similar arguments for the $\Delta_\ell$ 
and $\Delta_\ell^\prime$ variables. Instead of elaborating further, we 
present our naive expectations for all the ratio-type variables 
concisely in Table~2 below.

\begin{center}
\begin{tabular}{|c|cccc|}
\hline
Variable             & cMSSM & UED(5) & SM(4) & LH(T) \\
\hline\hline
$D_3$                &   M   &   M    &  L    &  S    \\[1mm]
$D_4$                &   M   &   M    &  S    &  S    \\[1mm]
$\widetilde{D}_3$    &   M   &   S    &  M    &  S    \\[1mm]
$\widetilde{D}_4$    &   M   &   S    &  M    &  S    \\[1mm]
$\Delta_\ell$        &   M   &   S    &  M    &  M    \\[1mm]
$\Delta_\ell^\prime$ &   M   &   S    &  M    &  --    \\[1mm]
\hline
\end{tabular}
\end{center}
\noindent
Table 2: {\footnotesize\it Naive expectations for ratio-type 
discriminators in the four models under consideration. The symbols used 
are to be read as follows. {\rm L} implies large, i.e. order 10 or more, 
{\rm M} implies medium, i.e. order unity or thereabouts, and {\rm S} 
implies small, i.e. order 0.1 or less. The empty slot arises because 
there are naively no hard jets to speak of in the LH(T) model and hence 
the ratio in question must depend on effects which cannot be predicted 
without carrying out a simulation.}

\bigskip\noindent We can now easily discriminate between these models by 
combining these ratio variables appropriately. For example, if we wish 
to discriminate between the cMSSM and UED(5) models, we consider 
$\widetilde{D}_{3,4}$ and the $\Delta_\ell, \Delta_\ell^\prime$. If we 
wish to discriminate between the cMSSM and the SM(4), we consider the 
$D_{3,4}$ and $\widetilde{D}_{3,4}$, and so on. A systematic study, 
considering the accessible parameter space in each of the four models is 
described in the next section.

\section{Addressing the Inverse Problem}

A statement of the full LHC Inverse Problem would be somewhat as 
follows: {\it If we observe a deviation from the SM predictions at the 
LHC, how do we determine its underlying cause? If it is due to new 
physics, how do we identify the model and determine its parameters?} 
This apparently-simple question actually requires one to cover a vast 
canvas of facts and inferences, as LHC data will be somewhat limited in 
scope. After all, the only observables will be a bunch of leptons and 
jets and their momenta in the transverse plane, and everything else will 
have to be inferred from this data.

\bigskip\noindent In this article, we concentrate on a limited portion 
of the general inverse problem. Through the previous sections, we have 
described the signals with $3\ell$ + MET + jets and $4\ell$ + MET + jets 
in $pp$ collisions at the LHC, which have little or no SM background. If 
a signal is seen in any of these two channels, four of the most likely 
new physics candidates are the models denoted cMSSM, UED(5), SM(4) and 
LH(T). The inverse problem in this case consists of trying to identify 
which of {\it these four} models is the cause of the observed signal. Of 
course, the signal, if seen, could also be due to some other model -- 
even, in principle, to completely new physics\footnote{Though that would 
come as a surprise, and is not, by its very definition, predictable.}.

\bigskip\noindent In the previous section, we have shown how we can 
define discriminating variables of two types, viz. event-counting type 
and ratio-type. The method is economical in the sense that the ratios 
are of pairs of event-counting variables and do not need separate data. 
In fact, the raw experimental data would require a trigger on leptons 
and MET, and the jet multiplicity for each event would be stored. This 
is a simple requirement and is certainly on the agenda for both the CMS 
and ATLAS Collaborations when the LHC becomes operational. With this 
limited data, we can construct all the variables defined in the previous 
section. At the present juncture, however, we do not have data, and in 
any case, we require to see how accurate are our guesses regarding these 
variables, as exhibited in Table~2. This requires a detailed Monte Carlo 
simulation of the events, which has been performed and is described 
below.

\bigskip\noindent The principal tools in our numerical analysis of the 
problem are
\begin{enumerate}
\item The event generator CalcHEP\cite{calchep}, which generates 
parton-level events for new physics models,
\item The event generator PYTHIA\cite{pythia}, which hadronises partons 
and creates hadron showers, collecting them into jets using a simple toy 
calorimeter algorithm called PYCELL, and, finally,
\item The software SUSPECT\cite{suspect} which creates a cMSSM spectrum 
from the standard input parameters and imposes known experimental 
constraints on it.
\end{enumerate}
In general, experimental constraints have been imposed in a very 
conservative fashion on all the models in question. For example, though 
three of the models have dark matter candidates, we have not imposed a 
constraint from the cosmic relic density\cite{DarkMatter}. Similarly, 
the LH(T) model might be severely constrained by low-energy 
data\cite{LHTlowenergy}, but this has not been explored in great detail 
yet. The philosophy adopted in this work is that we shall consider the 
maximum possible parameter ranges and see if we can avoid overlaps 
between the discriminating variables for variation over these full 
ranges. Further constraints can only cause the allowed parameter space 
--- and hence the range of the discriminating variables -- to shrink 
further, thereby {\it reducing} potential overlaps. What we exhibit in 
this article is, therefore, the worst case scenario in every variable. 
If a positive result is obtained notwithstanding this, then we may be 
sure that it can only be further strengthened by further experimental 
constraints\footnote{Unless, indeed, these are so strong as to rule out 
the model altogether}.

\bigskip\noindent To get a generic trigger which will work for each of 
the models under consideration, we have imposed the following kinematic 
cuts on leptons for identification:
\begin{itemize}
\item The transverse momentum $p_T^\ell$ should satisfy $p_T^\ell \geq 
10$~GeV. This is an acceptance cut required to excite showers in the 
electromagnetic calorimeter (EMC) for both LHC detectors.
\item The pseudorapidity $\eta_\ell$ should satisfy $\eta_\ell \leq 
2.5$. This roughly covers the barrel and end caps for the EMCs.
\item There should be no hadronic clusters with total energy $E_h > 0.3 
\ E_\ell$ in a cone of semi-vertical angle $\Delta R = 3.5$ around the 
lepton momentum as axis. This is an isolation criterion.
\end{itemize}
The criterion for labelling a lepton as `hard' has already been 
described in the context of the definitions in Table~1. In addition to 
this we require the total missing transverse momentum $p_T^{\rm miss}$ 
to satisfy
\begin{equation}
p_T^{\rm miss} \geq 20~~{\rm GeV} \ . 
\end{equation}

\noindent Once an event passes the above triggers, the hadronic final 
states are collected into jets. In our Monte Carlo simulations, we have 
used, as mentioned before, the PYCELL toy calorimeter routine.  To be 
identified as a jet, we impose the simple criteria:
\begin{itemize}
\item The transverse momentum $p_T^{\rm jet}$ should satisfy $p_T^{\rm 
jet} \geq 20$~GeV. This is an acceptance cut required to excite showers 
in the hadron calorimeter (HCAL) for both LHC detectors.
\item The pseudorapidity $\eta_{\rm jet}$ should satisfy $\eta_{\rm jet} 
\leq 2.5$. This roughly covers the barrel and end caps for the HCALs.
\end{itemize}
Energy and angular smearing are inbuilt in the PYCELL algorithm and are 
not imposed separately. Thus, we expect our simulated jets to resemble 
the actual jets observed at the LHC fairly closely. Naturally, in the 
LHC experiments, more sophisticated jet-identification techniques will 
be employed, but we do not expect to see dramatic deviations from our 
results. Thus, for each of the events simulated, we note the jet 
multiplicity, and this enables us, for a given luminosity, to generate 
the event-counting variables defined in Table~1. As for leptons, the 
criterion for labelling a jet as `hard' has already been described. Once 
we have all this information, it is a simple matter to compute the 
ratios defined in Equation~(\ref{eqn:Discriminators}), subject to a 
requirement that both numerator and denominator should be greater than 
5, for the given luminosity.
 
\bigskip\noindent In order to carry out a Monte Carlo simulation, we 
have generated 30,000 pair-production events for each point in the 
parameter space for each of the models under consideration. This large 
number corresponds, for a cross-section of 1~pb, to a luminosity of 
30~fb$^{-1}$, which is a conservative estimate of the maximum attainable 
at the LHC. Of course, for lower luminosities, all event-counting 
variables simple scale as the luminosity, unless, indeed, the numbers 
are very small and sensitive to statistical fluctuations.

\bigskip\noindent A detailed scan over the allowed parameter space for 
each of the four models in question would require an immense amount of 
computer time, and is probably not called for at the present juncture. 
In any case, our purpose in this article is merely to illustrate how, by 
correlating discriminator variables as defined, we can distinguish 
between models. To obtain a quick estimate of the spread in these 
variables as we scan the parameter space, we employ a standard trick, 
which is to make a {\rm random} sampling of the parameter space over its 
full accessible range. If a sufficiently large number of such random 
points is chosen, we expect a reasonable complete sampling of the 
parameter space, taking care of most of the kinematic as well as dynamic 
behaviour of the particles in the model. The possibility that there 
exists an isolated, un-sampled point in the parameter space where this 
behaviour would be very different is militated against by the fairly 
continuous behaviour of the spectrum and coupling constants in the 
models under question. The only singular behaviour really expected 
arises from particle resonances, but these are nicely smoothed out 
everywhere using the Breit-Wigner approximation. Thus we can argue that 
our random sampling of the parameter space actually leads to a fairly 
good representation of the spread in discriminating variables induced by 
scanning the parameter space. Before checking on these, however, it is 
necessary to mention the actual ranges over which the parameters are 
varied and the experimental constraints thereon. \begin{enumerate} \item 
In the cMSSM, the parameters are chosen randomly within the ranges given 
below: 
\vskip -15pt 
$$ 
\begin{array}{rcccl} 
100~{\rm GeV} & \leq & m_0 & \leq & 2~{\rm TeV} \\ 
100~{\rm GeV} & \leq & m_{1/2} & \leq & 2~{\rm TeV} \\ 
-2~{\rm TeV} & \leq & A_0 & \leq & +2~{\rm TeV} \\ 
5 & \leq & \tan \beta & \leq & 50 \\ 
{\rm sgn}(\mu) & = & +1 & {\rm or} & -1
\end{array} 
$$ 
The lower bounds on $m_0$ and $m_{1/2}$ roughly correspond to the bounds 
from LEP-2 and their upper bounds represent a stage when the squarks and 
gluinos are too heavy to be produced in any useful numbers at the LHC. 
The latter argument also applies to the magnitude of $A_0$. The range in 
$\tan \beta$ starts from the LEP and Tevatron lower bound and is cut off 
roughly at the point where the top quark Yukawa coupling to the charged 
Higgs bosons would become non-perturbative. The random parameter choice 
within these limits is fed into the software SUSPECT, which generates 
the cMSSM spectrum and couplings by solving the renormalization group 
(RG) equations evaluated at the one- and two-loop levels (as 
appropriate). Seven standard constraints\cite{SUSYbounds} are applied to 
the cMSSM parameter space as follows: \begin{enumerate} \item The 
electroweak parameter $\rho$ should satisfy $(\Delta\rho)_{\rm cMSSM} < 
2.2\times 10^{-3}$. \item The muon should satisfy $-7.7\times 10^{-10} < 
(g - 2)_\mu < 4.7\times 10^{-9}$. \item Radiative $b$-decays should 
satisfy $2.65\times 10^{-4} < {\rm BR}(B \to K^* \gamma) < 4.45 \times 
10^{-4}$. \item The lightest scalar $h^0$ should satisfy $M_h > 93$~GeV. 
\item The light chargino $\widetilde{\chi}_1^\pm$ should satisfy 
$M_{\widetilde{\chi}_1^\pm}> 104.5$~GeV. \item The light stop 
$\widetilde{t}_1$ should satisfy $M_{\widetilde{t}_1} > 101.5$~GeV. 
\item The light stau $\widetilde{\tau}_1$ should satisfy 
$M_{\widetilde{\tau}_1} > 98.8$~GeV. \end{enumerate} A further 
constraint on the parameter space arises from the requirement that it 
should be accessible at the LHC. Noting that a trilepton signal will 
call for at least a suppression by the leptonic\footnote{Recall that at 
the LHC we trigger only on electrons and muons, and not on taus.} 
branching ratios of a $W$ and a $Z$, i.e. $0.21 \times 0.067 = 0.014$, 
it is clear that unless the initial cross-section for pair production of 
squarks and gluinos is at least 10~fb, we will not get any signal worth 
analysing for 30~fb$^{-1}$ of data. Obviously, this constraint can be 
applied only after calculation of the cross-sections in question. 
Fortunately, this is mostly taken care of by the upper limits chosen for 
$m_0$ and $m_{1/2}$. Note that we have not applied the dark matter 
constraint which would require the observed cosmic relic density to 
match with its cMSSM prediction\cite{DarkMatter}. This, as explained 
before, is a conservative standpoint, and imposition of the dark matter 
constraint can only shrink the cMSSM parameter space further.

\bigskip\noindent Having decided on a viable point in the parameter 
space, then, the corresponding spectrum and couplings are generated by 
the SUSPECT codes and read in by the PYTHIA software, which is geared to 
`generate' gluino and squark pairs and simulate their cascade decays. 
Only those `events' are retained where the final state satisfies the 
10~fb constraint mentioned above and contains 3 or 4 identifiable 
leptons as well as missing $p_T$, as per the criteria given above. For 
these events, the jet multiplicity is counted. This enables us to 
classify the events into appropriate bins, whose accumulated numbers 
enable us to calculate the variables defined in Table~1 and 
Equation~(\ref{eqn:Discriminators}).

\item For the UED(5) model, there are only two parameters, viz. the size 
$R$ of the extra dimensions (usually represented by its inverse 
$R^{-1}$, and the cutoff $\Lambda$ for the theory, usually scaled as 
$\Lambda R$. We have chosen the points randomly in the ranges
$$
300~{\rm GeV} \leq R^{-1} \leq 1.5~{\rm TeV} 
\qquad\qquad\qquad
5 \leq \Lambda R \leq 20
$$
which are fairly conservative bounds for the UED(5) model. The lower 
bound\cite{UED5bounds} of 300~GeV on $R^{-1}$ is obtained by stretching 
the Tevatron constraint to its $5\sigma$ limit, while the upper bound of 
1.5~TeV is an accessibility limit as it more or less corresponds to 
cross-sections for $q_1$ and $g_1$ pair-production below the 10~fb limit 
explained above. The requirement that $\Lambda$ should remain above the 
LHC-accessible region determines the lower bound of 5 on $\Lambda R$, 
while the upper bound is determined by the requirement that the cutoff 
should not create a new and unnatural hierarchy of scales in the UED(5) 
model. Of course, both the upper and lower bounds 5 and 20 are ballpark 
numbers, which we use to estimate the parameter spread -- they could 
very well have been replaced by 4 and 22, for example (but not by, say, 
2 and 30).

\bigskip\noindent Once the point in the UED(5) parameter space is 
chosen, the couplings and parton-level cross-sections for production of 
$q_1q_1$, $q_1g_1$ and $g_1g_1$ pairs are calculated using the event 
generator CalcHEP, which has the provision for inclusion of new physics 
beyond the Standard Model. The parton-level momenta are fed into PYTHIA, 
using the LHEF formatting\cite{LesHouches}, and further development of 
hadronic final states is carried out by PYTHIA. The triggering, counting 
of jet multiplicity and calculation of discriminating variables then 
proceeds exactly as in the cMSSM case.

\item The analysis for the LH(T) model is done very similarly to the 
analysis in the UED(5) case. Here there are three parameters, viz, the 
symmetry-breaking scale $f$ and the Yukawa couplings $\kappa_q$ and 
$\kappa_\ell$. The ranges over which these are permitted to vary are
$$
500~{\rm GeV} \leq f \leq 1.5~{\rm TeV} \ ,
\qquad\qquad
0.25 \leq \kappa_q \leq 1 \ ,
\qquad\qquad
0.25 \leq \kappa_\ell \leq 1 \ .
$$
The lower bound on $f$ is induced by the lower bound $M(B_T) > 80$~GeV 
which arises from precision electroweak measurements\cite{LHTbounds}. 
The upper bound is an accessibility limit, since it more or less 
corresponds, as in the UED(5) case for $R^{-1}$, to a situation where 
the cross-section for $q_T$ pair-production falls below the 10~fb limit. 
Of course, every random choice of parameters in the LH(T) parameter 
space described above does not lead to trilepton and four-lepton signals 
at the LHC. The masses must be compatible with 
Equation~(\ref{eqn:Hierarchy}) in order to see such signals. Only those 
points which lead to such a mass hierarchy are, therefore, selected out 
of a large random sample. Once this has been done, the event generation 
can be carried out by a CalcHEP-PYTHIA combination through an LHEF 
interface\cite{Belyaev}, as in the case of the UED(5) model.

\item The simplest parameter space is the case of the SM(4), where we 
simply allow $M(b_4)$ to vary between 350~GeV and 600~GeV, i.e. between 
the experimental lower bound and the theoretical upper bound where the 
corresponding Yukawa coupling would lead to a Landau 
pole\cite{SM4facts}. The mass of the $t_4$, the other free parameter of 
the theory, is not relevant for our analysis, as has been explained 
before. The simulation of events in the SM(4) is done using PYTHIA, more 
or less in the same way as the previous cases.

\end{enumerate}  

\begin{figure}[htb]
\centerline{ \epsfxsize= 6.5 in \epsfysize= 2.6 in \epsfbox{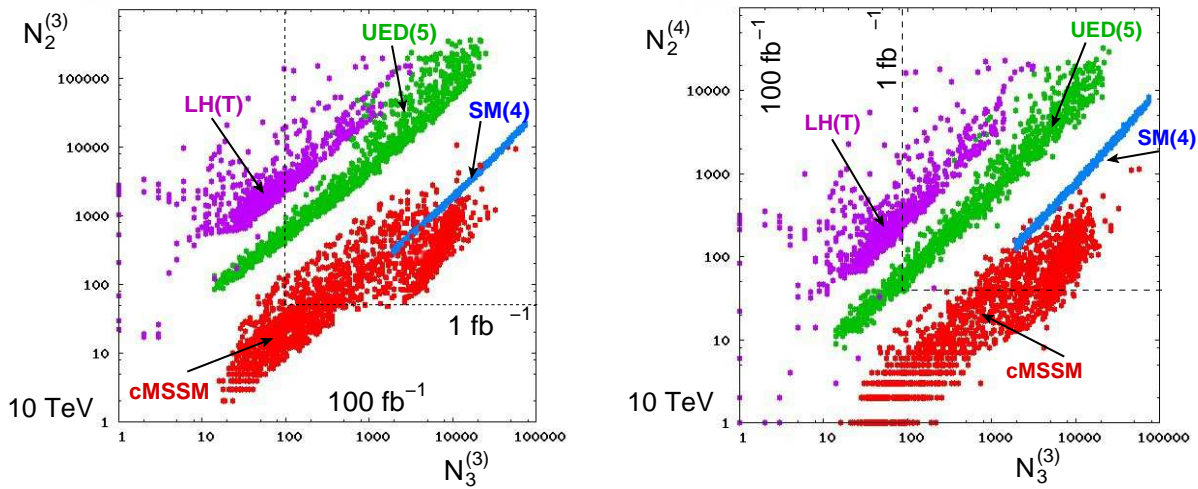} }
\caption{{\footnotesize\it Correlation plots between event counting-type 
variables ($a$) $N_2^{(3)}$ against $N_3^{(3)}$, and ($b$) $N_2^{(4)}$ 
against $N_3^{(3)}$. The box drawn with broken lines shows the reach 
with a luminosity of 1~fb$^{-1}$, for which the axis labels should be 
scaled by a factor of 0.01 on either axis. These numbers all correspond 
to $\sqrt{s} = 10$~TeV.}}
\label{fig:N33-N23-N24}
\end{figure}
\vskip -5pt

\noindent We are now in a position to exhibit and discuss our results. 
The simplest way to discrimininate between models is to correlate a pair 
of event counting-type variables. Several of these are defined in 
Table~1, and can be plotted against each other. The best results arise 
when we plot ($a$) $N_2^{(3)}$ against $N_3^{(3)}$, and ($b$) 
$N_2^{(4)}$ against $N_3^{(3)}$. These are exhibited in 
Figure~\ref{fig:N33-N23-N24}, where the left box corresponds to the 
option ($a$) and the right box to the option ($b$). This correlation 
plot is a two-dimensional example of a `signature space' defined in 
Ref.~\cite{LHCinverse}. We may note that the variable $D_3$ is the slope 
of the graph in the box ($a$).

\noindent There are two nested boxes in Figure~\ref{fig:N33-N23-N24}. 
The outer one, which corresponds to a rather optimistic projection of an 
integrated luminosity of 100~fb$^{-1}$, is correctly labelled on the two 
axes. The inner one, two sides of which are represented by broken lines, 
indicates what would be obtained with an integrated luminosity of 
1~fb$^{-1}$, i.e. an early result. For the 1~fb$^{-1}$-option, the 
labels on the two axes must be scaled by a factor of $\frac{1}{100}$. 
Within the boxes, clusters of points are colour-coded to represent the 
different models, but are also labelled alongside and indicated with 
arrows to avoid confusion in a black-and-white printout. Each point of a 
given colour on the graph represents a point in the parameter space. The 
clusters of points of a given colour represent the Monte Carlo 
equivalent of a filled region of parameter space, and hence the rough 
boundaries of each cluster may be regarded as an approximate contour 
delineating the region predicted by a specific model on the plot in 
question.

\bigskip\noindent What are the conclusions one can draw from 
Figure~\ref{fig:N33-N23-N24}? If one considers the left box, where 
$N_2^{(3)}$ is plotted against $N_3^{(3)}$, it is clear that there are 
two distinct regions -- one which is populated by the 
slightly-overlapping UED(5) and LH(T) clusters, and one which is 
populated by the overlapping cMSSM and SM(4) clusters. The distance 
between these two regions is not small if we note that this is a 
logarithmic plot. Thus, the qualitative results will not change even if 
we see fairly substantial corrections due to PDFs, radiative effects, 
etc. The reality, of course, should ideally correspond to a single point 
in this graph, but in practice, we will get a square blob due to 
experimental errors. There are now three possibilities, each calling for 
a separate comment.
\begin{itemize}
\item If the experimental blob should lie squarely in one of the regions 
where just one of the four models is unambiguously indicated, this will 
provide an immediate solution to the inverse problem for this signal. 
For example, if $N_3^{(3)} = 100\pm 10$ and $N_2^{(3)} = 10\pm 3$, the 
cMSSM must be the new physics, and the other three models are ruled out.
\item If the experimental blob lies in or near the overlap region of two 
of the models in question, one cannot distinguish between these two 
models (by looking only at this graph), but the other two models get 
ruled out. For example, if $N_3^{(3)} = 1000\pm 32$ and $N_2^{(3)} = 
20000\pm 140$, then the new physics may be either the UED(4) or LH(T) 
models, but the cMSSM and the SM(4) are definitely ruled out. The 
channel separating the cMSSM and the UED(5), which separates the 
clusters into two distinct parts, is sufficiently wide so that even with 
an error $\propto \sqrt{N}$, an experimental blob lying close to one of 
the regions cannot have an overlap with the other region. Thus, we may 
be sure that at least two of the models will always be eliminated.
\item If the experimental blob lies away from {\it all} the clusters, 
e.g. we have $N_3^{(3)} = 10000\pm 100$ and $N_2^{(3)} = 10\pm 3$, then 
we will have a signal for new physics, indeed, but it can correspond to 
none of the models under consideration. In that case, we would not have 
a solution for the inverse problem, but will have the somewhat cold 
consolation that at least four of the most popular models are ruled out.
\end{itemize}
The purposes of this work are served if we consider the first two 
possibilities; the third requires an altogether different kind of study. 
If the first should happen, we can consider ourselves lucky and treat 
the subsequent discussion as a method to shore up the conclusions 
already reached. If the second case should happen, which is equally 
probable, we now require to see how one can further disentangle the 
models by looking at other variables. The first thing to do is to look 
at other event-counting variables. The best combination is exhibited in 
the right box in Figure~\ref{fig:N33-N23-N24}, where an trileptonic 
variable and a four-leptonic variable are plotted against each other. As 
we have already discussed, four-lepton signals show more variation in 
the jet complement than trilepton signals, especially for the SM(4), and 
hence, a better discrimination may be obtained by using a four-leptonic 
variable. This, is, in fact, apparent when we look at the plot of 
$N_3^{(4)}$ against $N_3^{(3)}$. Not only is the channel between the 
cMSSM and UED(5) clusters wider and cleaner, but now the overlap between 
the UED(5) and LH(T) clusters is very small, and similarly, the narrow 
band representing the SM(4) has almost disengaged itself from the cMSSM 
cluster. The possibility of having an unambiguous verdict from the 
experimental data is, therefore, somewhat better in this plot than in 
the one on the left. However, it has to be admitted that complete 
disambiguation cannot be done by considering these two figures, either 
in isolation, or taken together.
 
\bigskip\noindent We take note of the fact that if the experimental 
signal is large enough to be seen even with 1~fb$^{-1}$ of data, the 
discrimination properties of the plots in Figure~\ref{fig:N33-N23-N24} 
(or the lack thereof) remain undiminished. Increase in the luminosity 
will make more of the parameter space accessible (except for the SM(4) 
case), but in separate patches outside the box with broken lines, where 
there is practically no overlap, except for a few outlying points.

\begin{figure}[htb]
\centerline{\epsfxsize= 6.3 in \epsfysize= 5.1 in \epsfbox{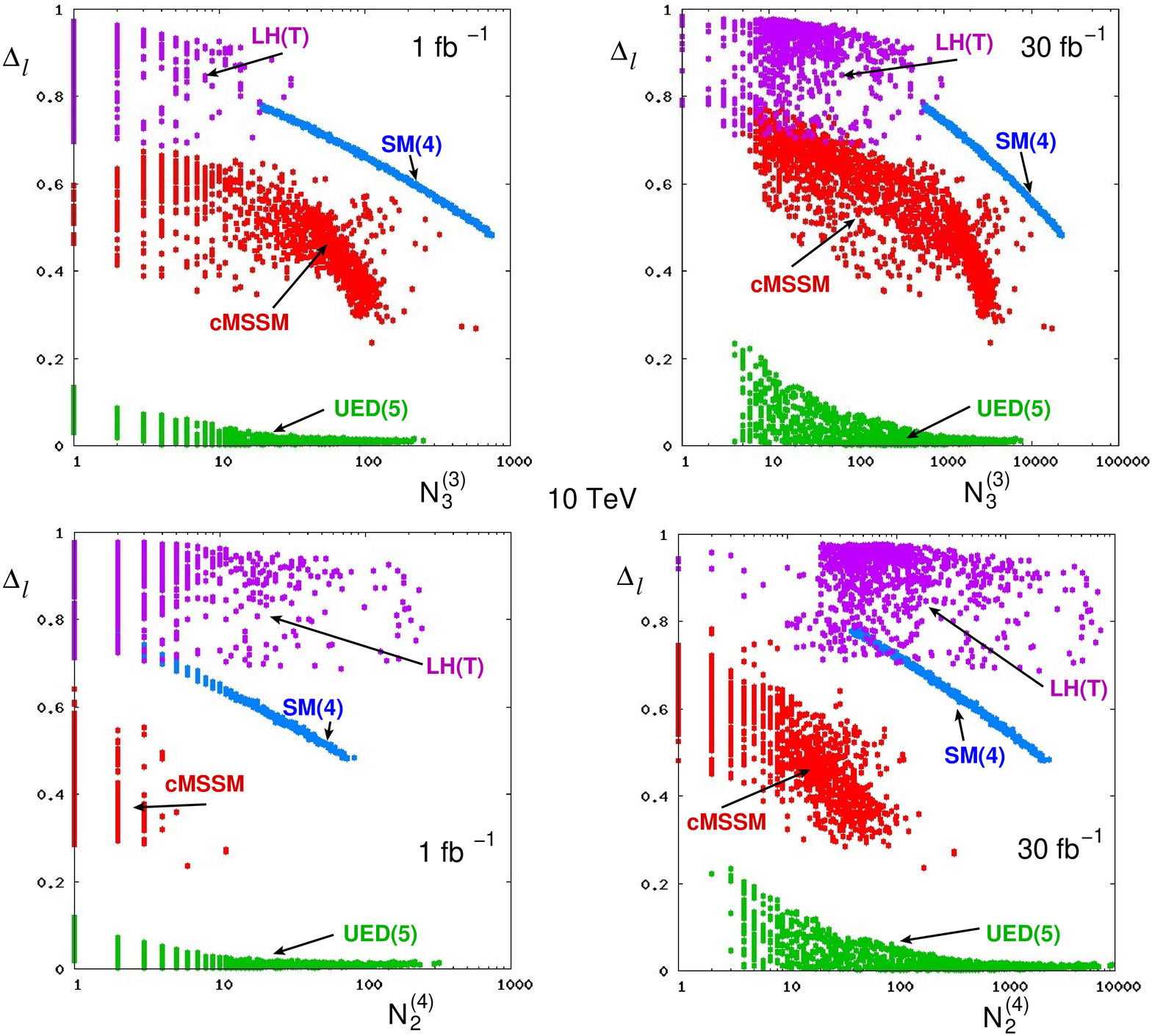} }
\caption{{\footnotesize\it Correlation plot between event counting-type 
variables versus a ratio-type variable for $\sqrt{s} = 10$~TeV. The 
upper boxes have the ratio $\Delta_\ell$ plotted against the number 
$N_3^{(3)}$, and the lower boxes show the same ratio $\Delta_\ell$ 
plotted against the number $N_2^{(4)}$. In each case, the box on the 
left (right) shows the prediction for ${\cal L} = 1~$~fb$^{-1}$ 
(30~fb$^{-1}$). }}
\label{fig:N33-N24-deltaL}
\end{figure}

\bigskip\noindent The option of using only event counting-type variables 
having achieved partial, but not complete, success indicates that we 
should next try to combine event counting-type variables with ratio-type 
variables to get a better discriminatory power. Here the best options 
are the plots shown in Figure~\ref{fig:N33-N24-deltaL}, where the upper 
boxes have the ratio $\Delta_\ell$ plotted against the number 
$N_3^{(3)}$, and the lower boxes show the same ratio $\Delta_\ell$ 
plotted against the number $N_2^{(4)}$. In each case, the box on the 
left shows the prediction for a luminosity of 1~fb$^{-1}$, i.e. the 
early data at the LHC, while the box on the right corresponds to 
30~fb$^{-1}$, which may well be the final data sample at the LHC.  Note 
that there is no real qualitative differences between the two luminosity 
options, except that in the high luminosity case, many more points in 
the parameter space are accessible, which is as expected.
 
\vskip -10pt

\bigskip\noindent If we recall that the problem in 
Figure~\ref{fig:N33-N23-N24} was that of separating the cMSSM-SM(4) and 
UED(5)-LH(T) cluster pairs, then Figure~\ref{fig:N33-N24-deltaL} is very 
interesting. Obviously, the UED(5) and LH(T) clusters are now widely 
separated, with $\Delta_\ell$ playing the role of a single-variable 
discriminator. If we look at the definition of $\Delta_\ell$ in 
Equation~(\ref{eqn:Discriminators}) and Table~1, it is clear that this 
is simply related to the fact that in the LH(T) model we will almost 
always have a hard lepton, while in the UED(5) model, that is a rarity. 
At the same time, there is also a clear separation of the SM(4) cluster 
from the cMSSM cluster, except for some outlying cMSSM points which 
appear for the higher luminosity. Even so, the separation between these 
clusters in the plot in the lower boxes is quite substantial -- this is 
not unexpected, since the abscissa is a four-lepton variable, where the 
SM(4) has a clear bias towards a smaller number of jets. Thus, if we 
combine the information from Figure~\ref{fig:N33-N23-N24} as well as 
Figure~\ref{fig:N33-N24-deltaL}, it is clear that we will know quite 
clearly which of the four models under study is the cause of an excess 
leptonic signal.

\begin{figure}[ht]
\centerline{\epsfxsize= 6.5 in \epsfysize= 5.3 in \epsfbox{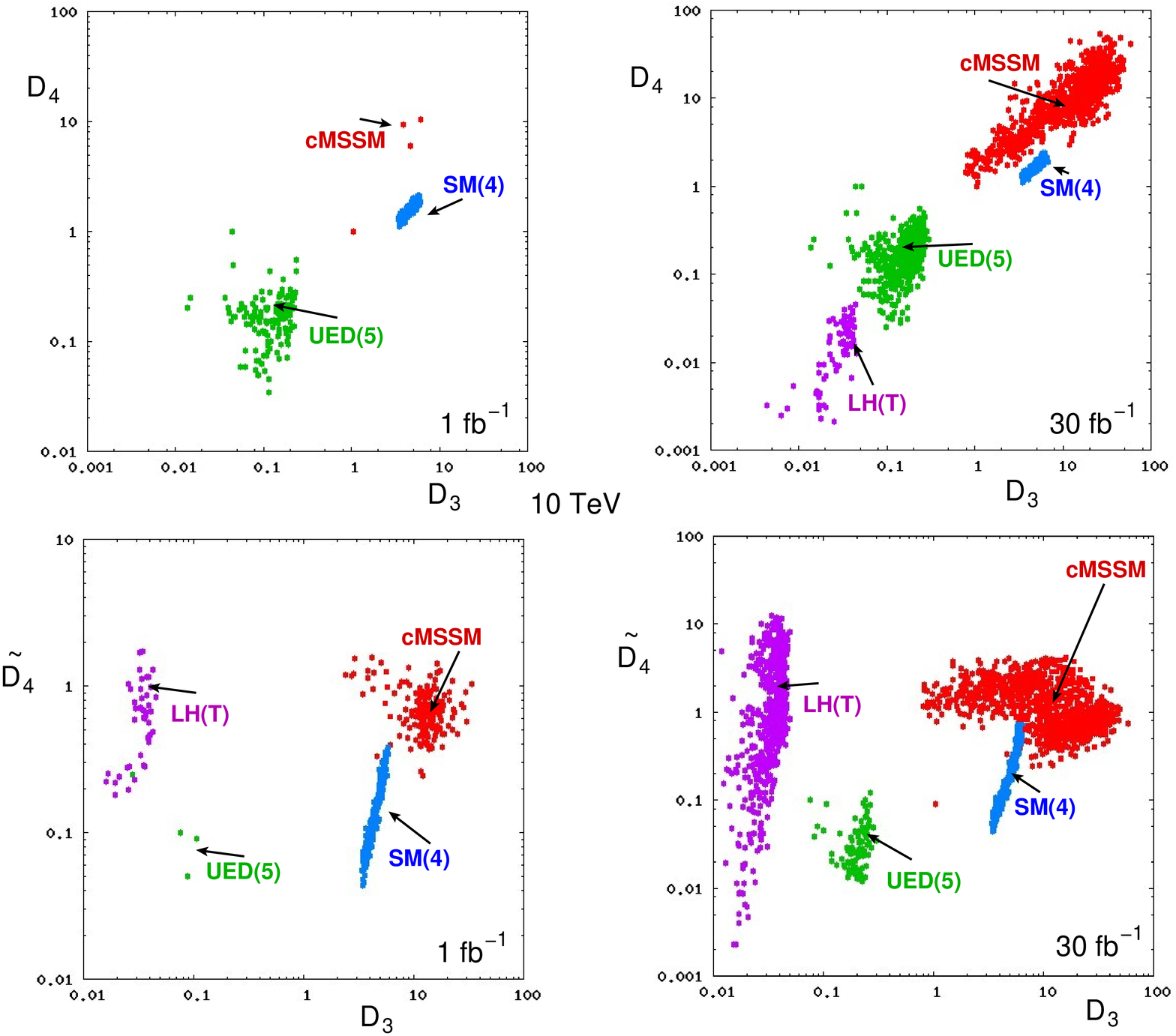} }
\caption{{\footnotesize\it Correlation plot between ratio-type variables 
for $\sqrt{s} = 10$~TeV. The upper boxes have the ratio $D_4$ plotted 
against $D_3$, and the lower boxes show the ratio $\widetilde{D}_4$ 
plotted against the same variable $D_3$. In each case, the box on the 
left (right) shows the prediction for ${\cal L} = 1~$~fb$^{-1}$ 
(30~fb$^{-1}$).}}
\label{fig:D3-D4-DT4}
\end{figure}
\vskip -5pt

\noindent Our analysis could have stopped at this point, since the goal 
of disentangling signals for each of the four models in question has 
been achieved. However, we have even nicer results when we correlate 
pairs of ratio-type variables as defined in 
Equation~(\ref{eqn:Discriminators}). In Figure~\ref{fig:D3-D4-DT4} we 
plot the ratios $D_3$ and $D_4$ against each other in the upper half, 
and the ratios $D_3$ and $\widetilde{D}_4$ in the lower half, for 
luminosities of 1~fb$^{-1}$ and 30~fb$^{-1}$ on the left and right 
respectively. A glance at the figure will show that we have four clearly 
delineated clusters of points corresponding to the four models in 
question. We note that these are logarithmic plots, and hence, what 
looks like a small separation between, say the cMSSM cluster and the 
SM(4) cluster, is actually quite significant. Thus, if signals are seen 
at the lower luminosity of 1~fb$^{-1}$, clearly, any one of the plots 
should achieve, at one go, a clear disambiguation between all four 
models. If signals appear later, when 30~fb$^{-1}$ have been collected, 
there may be some difficulty in separating the SM(4) from the cMSSM, as 
the clusters are quite close. We have already discussed why the clusters 
grow in size when the luminosity increases.

\bigskip\noindent We can make several plots of similar nature, for 
example, by taking the four-lepton variable $D_4$ and plotting it 
severally against $\Delta_\ell$ and $\Delta_\ell^\prime$. Most of these 
plots show the following common features:
\begin{itemize}
\item Each of the models results in a separate cluster of points, which 
is somewhat scattered for 1~fb$^{-1}$ but compact and dense for 
30~fb$^{-1}$.
\item In each case, there is a small overlap or proximity between just 
one pair of the models in question, others being widely separated.
\item In general, the hardest to separate seem to be the cMSSM cluster 
and the SM(4) cluster.
\end{itemize}

\begin{figure}[h]
\centerline{\epsfxsize= 6.5 in \epsfysize= 5.3 in \epsfbox{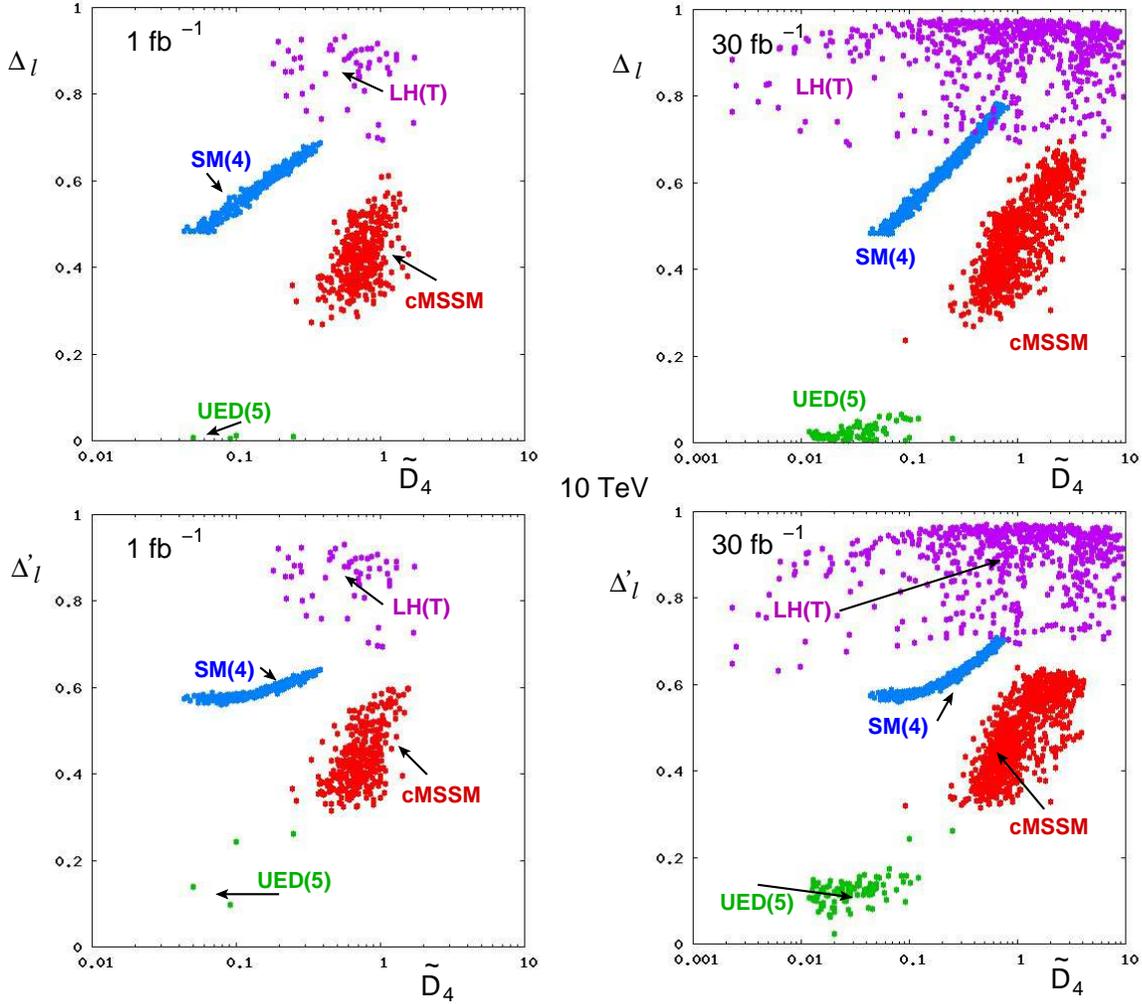} }
\caption{{\footnotesize\it Correlation plot between ratio-type variables 
for $\sqrt{s} = 10$~TeV. The upper boxes have the ratio $D_4$ plotted 
against $\Delta_\ell$, and the lower boxes show the same ratio 
$\widetilde{D}_4$ plotted against $\Delta_\ell^\prime$. As before, the 
box on the left (right) shows the prediction for ${\cal L} = 
1~$~fb$^{-1}$ (30~fb$^{-1}$) in each case.}}
\label{fig:DT4-deltaL-deltaLp}
\end{figure}

\noindent The last problem is not insurmountable, however, as we have 
already seen in the context of Figure~\ref{fig:N33-N24-deltaL}. This 
finds its neatest manifestation in Figure~\ref{fig:DT4-deltaL-deltaLp}, 
where we plot $\widetilde{D}_4$ against $\Delta_\ell$ and 
$\Delta_\ell^\prime$ respectively. A glance at 
Figure~\ref{fig:DT4-deltaL-deltaLp} will show that there is now a clear 
separation between the (blue) SM(4) cluster and the (red) cMSSM cluster. 
Experimental errors will be too small to permit any experimental blob to 
span this gulf. Thus, even if all else fails, this plot will be enough 
to distinguish between the SM(4) and cMSSM cases. The plot is also good 
enough to distinguish the UED(5) uniquely, since the corresponding 
points form another isolated cluster. What it fails to do is to 
completely separate the SM(4) from the LH(T) model, but that is not a 
cause of worry, since these two models lead to widely separated clusters 
(by more than two orders of magnitude in $D_3$) in 
Figure~\ref{fig:D3-D4-DT4}.

\bigskip\noindent We can thus present a clear and unambiguous 
prescription for solving the LHC inverse problem in trilepton and 
four-lepton signals. All that requires to be done is to
\begin{enumerate}
\item trigger on events with $3\ell$ + MET with or without accompanying 
jets, as well as on events with $4\ell$ + MET with or without 
accompanying jets;
\item using a suitable jet counting algorithm, determine the jet 
multiplicity for each of the triggered events, and use this information 
to populate jet-multiplicity distributions for both kinds of signal;
\item calculate the event counting-type variables of Table~1, using the 
jet-multiplicity distributions and simple variables like lepton and jet 
$p_T$;
\item calculate the ratio-type variables defined in 
Equation~(\ref{eqn:Discriminators});
\item make correlation plots as shown in this article: the solution to 
the inverse problem will leap out of the figure.
\end{enumerate}

\vskip -5pt

\noindent Before concluding this section, it is relevant to say that the 
above prescription does not require any real extra effort on the part of 
our experimentalist colleagues over and above what they would be doing 
anyway when the LHC data arrives. Trilepton and four-lepton signals with 
MET are at the forefront of the new physics programme at the LHC in any 
case, and it should require a small modification of the usual analysis 
to achieve the results projected in the present work.

\section{Critical Summary}

\noindent This work has grown out of the oft-quoted statement that 
trilepton and four-lepton signals (accompanied by MET and jets) could be 
like `smoking gun' signals of supersymmetry when sufficient data are 
accumulated at the LHC. Such statements beg the retort that the same 
signals could be produced by other models as well, and in this paper, we 
have considered three popular alternatives, viz, the UED(5) or `bosonic 
supersymmetry', the LH(T) and the simplest extension of the SM, which 
involves adding a fourth sequential generation of fermions. We have 
explained at length how this happens, what are the decay chains 
responsible and what kind of parameters one requires to have a 
competitor for supersymmetry, embodied as the cMSSM. Our analysis has 
been carried out as objectively as possible, i.e. without any bias in 
favour of or against supersymmetry or any other model.

\bigskip\noindent The major points made in the article are as follows. 
We first note that at the LHC, the most viable signatures for new 
physics beyond the SM will occur when we have heavy particles produced 
through strong interactions, but decaying into leptonic final states. 
This can happen only if there is a conserved (or nearly conserved) 
quantum number), and three of the four models (cMSSM, UED(5) and LH(T)) 
do have such a quantum number corresponding to a discrete $Z_2$ symmetry 
in the theory. This also makes the lightest particle carrying a 
nontrivial $Z_2$ quantum number a stable particle and hence a dark 
matter candidate. Thus models of this nature are important to 
investigate. We focus on final states with 3 or 4 leptons, missing 
transverse momentum and an unspecified number of jets. These are shown 
(Figure~\ref{fig:Cascades}) to arise in all of the models in question, 
which brings us to an inverse problem -- if we do see a signal of this 
nature, how do we know which of the models is the correct one? We then 
show (Figure~\ref{fig:Distributions}) that the lepton $p_T$ is not 
enough to distinguish between these models, because it can look 
identical for certain parameter choices of the models. However, for the 
same choices, the jet multiplicity distributions look very different. 
Using this hint and some general arguments based on the event topology 
in the individual models, we then go on to define a set of 
discriminating variables (Table~1 and 
Equation~(\ref{eqn:Discriminators})). Finally we do a numerical survey 
of the models in question, creating a cluster of points in each 
discriminating variable by random sampling of points in the parameter 
space. It is seen that by plotting suitable pairs of the discriminating 
variables against each other 
(Figures~\ref{fig:N33-N23-N24}--\ref{fig:DT4-deltaL-deltaLp}), the four 
models can generally be made to fall into distinct regions of the 
diagram, with some small degree of overlap. The latter is not a serious 
problem, because models which show overlap for one pair of variables get 
widely separated for a different set of variables. Hence, we arrive at a 
rather elegant way of solving the LHC inverse problem in trilepton and 
four-lepton signals (accompanied by MET and jets) -- there will be only 
one point (with error bars) in each of these planes when the 
experimental data are available, and one only has to see where it lies 
to get a spectacular solution to the inverse problem.

\bigskip\noindent Our work and what it hopes to achieve is summed up in 
the previous paragraph. To put this in perspective, we now come to some 
of the issues which our work does {\it not} address. Apart from 
technical details, such as the quality of Monte Carlo simulation and jet 
reconstruction -- where we feel confident that our results, though not 
as refined as one might have wished, have not gone in the wrong 
direction -- there are two major scenarios where our work could prove 
inadequate.
\begin{enumerate}
\item Clearly, in order to identify a particular model as the new 
physics source, the experimental data should map, in each one of the 
Figures~\ref{fig:N33-N23-N24}--\ref{fig:DT4-deltaL-deltaLp}, to points 
which lie in the region corresponding to a particular model. Thus, if 
the LH(T) is the correct option, the experimental blob described in the 
last section should, in every case, lie within the (violet) cluster of 
points indicating the LH(T). Deviations from this can happen either 
($a$) when the experimental blob matches with the LH(T) in some 
diagrams, but not in others, or ($b$) when the experimental blob always 
lies outside the area of scatter corresponding to all the four models 
under consideration. Any deviation of this nature must be interpreted as 
due to a different kind of new physics. Even a single mismatch will 
indicate that new theoretical ideas (or at least some modification of 
these models) are required.
\item The requirement that the experimental blob should lie in the 
appropriate region of parameter space is a necessary condition to 
identify a certain model, but not a sufficient one. For example, if in 
all the diagrams, the data match with the cMSSM patches, then we may say 
that supersymmetry is strongly indicated, but it does not necessarily 
have to be the constrained version which we have been using in this 
analysis. Supersymmetric models which relax the universality assumptions 
of the cMSSM, or even the 124-parameter MSSM with no universality or 
unification assumptions, are not ruled out by this test alone.
\end{enumerate}
In fact, as mentioned in an earlier section, this article discusses four 
of the popular constructions beyond the minimal SM which could lead to 
trilepton and four-lepton signals at the LHC. There are other 
possibilities, which have not been touched upon in this paper. Though 
one can provide arguments against such options (for example, 
$R$-parity-violating supersymmetry is unpopular because it does not 
provide a dark matter candidate), Nature has surprised us before and may 
do so again. Thus, we can only claim to have set out a neat technique to 
approach the inverse problem. We hope that future studies of new physics 
models in the trilepton and four-lepton channels will find this a useful 
framework in which to present the corresponding results.

\bigskip\noindent Having mentioned the caveats and worst case scenarios, 
we now take a positive approach and see what will happen if, indeed, the 
experimental blobs correspond, in each diagram, to the cluster 
corresponding to a particular model. This will immediately rule out the 
three other models\footnote{This will, of course, apply only to the 
minimal version of these models. Exotic extensions of any model can 
probably be created {\it \'a posteriori} to spread the corresponding 
cluster to the experimental blob.}, but it will not tell us which part 
of the parameter space of the successful model is responsible for the 
signal. However, one can then do a fine-grained parameter scan 
(systematic or random), for the entire set of discriminating variables, 
and see which point(s) in the parameter space can fit the whole set of 
experimental data. This, rather computer-intensive, study can be done 
only when there is actual data, but it has the merit of determining all 
the parameters of the model together. Thus, if we can find a fitting 
point, the entire mass spectrum and the couplings are determined by the 
theoretical structure of the model.

\bigskip\noindent In conclusion, therefore, we have studied a particular 
aspect of the LHC inverse problem in a more intensive way than general 
studies of the inverse problem have done previously. We have devised a 
set of rather robust kinematic variables and set up a set of correlation 
plots which could lead to spectacular solutions of the inverse problem 
when the LHC data become available. Our work has the advantages of 
simplicity and economy, since we only suggest the use of data which will 
surely be collected by the ATLAS and CMS experiments, and we use only 
number-counting variables, which are more robust than others against 
corrections due to smearing and other effects. We, therefore, conclude 
this article with a hope that the techniques suggested here will find 
favour in the community as a simple and direct approach to the inverse 
problem at the LHC -- a problem which is likely to assume paramount 
importance in a year or two from the present.

\bigskip\bigskip\noindent {\it Acknowledgments}: {\footnotesize The 
authors gratefully acknowledge discussions with M. Guchait and G. 
Majumdar, both of the CMS Collaboration. SR would like to thank the 
Department of Physics, University of Calcutta, for hospitality while the 
problem was being formulated. BB and AK would like to acknowledge the 
hospitality of the Department of Theoretical Physics, Tata Institute of 
Fundamental Research, where most of this work was done. We would like to 
acknowledge partial financial support from the University Grants 
Commission, India (BB, AK), the Council for Scientific and Industrial 
Research, Government of India (AK), the Board of Research in Nuclear 
Sciences, Government of India (AK) and, finally, the Academy of Finland 
(SKR).}


\newpage
\begin{thebibliography}{99}
\bibitem{LHCwebsite}
Regular updates are available at the URL \\
http://lhc.web.cern.ch/lhc/News.htm

\bibitem{LHCdijetCS}
M.~Cardaci [ATLAS Collaboration and CMS Collaboration], 
arXiv:0805.2906 [hep-ex].

\bibitem{cMSSM}
For a review, see, for example, S.~P.~Martin, arXiv:hep-ph/9709356.
 
\bibitem{SUSYatLHC}
For a recent review, see, for example, N.~Ozturk, f.~t.~ATLAS and 
CMS~Collaborations, arXiv:0910.2964 [hep-ph];

\bibitem{UED5model}
T.~Appelquist, H.~C.~Cheng and B.~A.~Dobrescu, Phys.\ Rev.\  D {\bf 64}, 
035002 (2001)

\bibitem{UED5spectrum}
H.~Georgi, A.~K.~Grant and G.~Hailu, Phys.\ Lett.\  B {\bf 506}, 207 (2001) ; \\
H.~C.~Cheng, K.~T.~Matchev and M.~Schmaltz, Phys.\ Rev.\  D {\bf 66}, 036005 
(2002).

\bibitem{UED5signals}
T.~G.~Rizzo, Phys.\ Rev.\  D {\bf 64}, 095010 (2001); \\
H.~C.~Cheng, K.~T.~Matchev and M.~Schmaltz, Phys.\ Rev.\  D {\bf 66}, 
056006 (2002); \\
C.~Macesanu, C.~D.~McMullen and S.~Nandi, Phys.\ Rev.\  D {\bf 66}, 
015009 (2002); \\
C.~Macesanu, C.~D.~McMullen and S.~Nandi, Phys.\ Lett.\  B {\bf 546}, 
253 (2002); \\
H.~C.~Cheng, Int.\ J.\ Mod.\ Phys.\  A {\bf 18}, 2779 (2003); \\
A.~Muck, A.~Pilaftsis and R.~Ruckl, Nucl.\ Phys.\  B {\bf 687}, 55 (2004); \\
C.~Macesanu, S.~Nandi and M.~Rujoiu, Phys.\ Rev.\  D {\bf 71}, 036003 (2005); \\
G.~Bhattacharyya, et al, Phys.\ Lett.\  B {\bf 628}, 141 (2005); \\
A.~Datta and S.~K.~Rai, Int.\ J.\ Mod.\ Phys.\  A {\bf 23}, 519 (2008); \\
S.~K.~Rai, Int.\ J.\ Mod.\ Phys.\  A {\bf 23}, 823 (2008); \\
B.~Bhattacherjee, et al, Phys.\ Rev.\  D {\bf 78}, 115005 (2008); \\
G.~H.~Brooijmans {\it et al.}, arXiv:0802.3715 [hep-ph]; \\
G.~Bhattacharyya, et al, Nucl.\ Phys.\  B {\bf 821}, 48 (2009).

\bibitem{LittleHiggs}
N. Arkani-Hamed et al., JHEP {\bf 08}, 021 (2002); \\
N.~Arkani-Hamed, et al, JHEP {\bf 0207}, 
034 (2002).

\bibitem{LHTmodel}
H.C. Cheng and I. Low, JHEP {\bf 09}, 051 (2003) and JHEP {\bf 08}, 061 (2004).

\bibitem{LHTsignals}
M.~Perelstein, Prog.\ Part.\ Nucl.\ Phys.\  {\bf 58}, 247 (2007); \\
J.~Hubisz and P.~Meade, Phys.\ Rev.\  D {\bf 71}, 035016 (2005); \\
T.~Han, et al, Phys.\ Rev.\  D {\bf 67}, 095004 (2003); \\
A.~Belyaev, et al, Phys.\ Rev.\  D {\bf 74}, 115020 (2006); \\
M.~S.~Carena, et al, Phys.\ Rev.\  D {\bf 75}, 091701 (2007).

\bibitem{SM4facts}
B.~Holdom, et al, arXiv:0904.4698 [hep-ph].

\bibitem{DarkMatter}
For reviews, see: \\
G. Jungman, M. Kamionkowski and K. Griest, Phys. Rept. {\bf 267}, 195 (1996); \\
G. Bertone, D. Hooper and J. Silk, Phys. Rept. {\bf 405}, 279 (2005); \\ 
D.~Hooper and S.~Profumo, Phys.\ Rept.\  {\bf 453}, 29 (2007). \\
Also see: \\
J.~Lubish and P.~Meade, Phys.\ Rev.\ D {\bf 71}, 035016 (2005); \\
A.~Birkedal, et al, Phys.\ Rev.\ D {\bf 74}, 
035002 (2006).

\bibitem{LHCinverse}
N.~Arkani-Hamed, et al, JHEP {\bf 0608}, 070 (2006) ; \\
N.~Arkani-Hamed, et al, arXiv:hep-ph/0703088; \\
B.~C.~Allanach, et al, JHEP {\bf 0708}, 023 (2007); \\
J.~Hubisz, et al, Phys.\ Rev.\  D {\bf 78}, 075008 (2008); \\
A.~Datta, et al, Phys.\ Lett.\  B {\bf 659}, 308 (2008) ; \\
A.~Belyaev {\it et al.}, Pramana {\bf 72}, 229 (2009); \\
J.~L.~Kneur and N.~Sahoury, arXiv:0808.0144 [hep-ph]; \\
C.~F.~Berger, et al, JHEP {\bf 0902}, 023 (2009); \\
J.~J.~Heckman, et al, arXiv:0903.3609 [hep-ph]; \\
C.~Balazs and D.~Kahawala, arXiv:0904.0128 [hep-ph]; \\
S.~S.~AbdusSalam, et al, arXiv:0904.2548 [hep-ph]; \\
K.~T.~Matchev, F.~Moortgat, L.~Pape and M.~Park, JHEP {\bf 0908}, 104 (2009); \\
D.~E.~Lopez-Fogliani, et al, arXiv:0906.4911 [hep-ph]; \\
G.~Belanger, et al, arXiv:0906.5048 [hep-ph]; \\
L.~Roszkowski, R.~Ruiz de Austri and R.~Trotta, arXiv:0907.0594 [hep-ph]; \\
O.~Buchmueller {\it et al.}, arXiv:0907.5568 [hep-ph]; \\
B.~Altunkaynak, arXiv:0909.5246 [hep-ph].

\bibitem{SUSYleptons}
M.~Bisset, et al, JHEP {\bf 0908}, 037 (2009); \\
Z.~Sullivan and E.~L.~Berger, Phys.\ Rev.\  D {\bf 78}, 034030 (2008); \\
S.~Bhattacharya, A.~Datta and B.~Mukhopadhyaya, Phys.\ Rev.\  D {\bf 78}, 
115018 (2008) and JHEP {\bf 0710}, 080 (2007); \\
N.~Bhattacharyya and A.~Datta, Phys.\ Rev.\  D {\bf 80}, 055016 (2009).

\bibitem{pythia}
T.~Sjostrand, S.~Mrenna and P.~Skands, JHEP {\bf 0605}, 026 (2006).

\bibitem{calchep}
A.~Pukhov, arXiv:hep-ph/0412191.

\bibitem{CTEQ5}
H.~L.~Lai {\it et al.}  [CTEQ Collaboration], Eur.\ Phys.\ J.\  C {\bf 12}, 
375 (2000).

\bibitem{UED5bounds}
K.~Agashe, N.G.~Deshpande, and G.H.~Wu, \PL(B514,309,2001); \\
D.~Chakraverty, K.~Huitu, and A.~Kundu, \PL(B558,173,2003); \\
J.F.~Oliver, J.~Papavassiliou, and A.~Santamaria, \PR(D67,056002,2003); \\
A.J.~Buras, M.~Spranger, and A.~Weiler, \NP(B660,225,2003); \\
A.J.~Buras \etal, \NP(B678,455,2004); \\
U.~Haisch and A.~Weiler, \PR(D76,034014,2007); \\
I.~Gogoladze and C.~Macesanu, \PR(D74,093012,2006); \\ 
T.~Fl\"acke, D.~Hooper and J.~March-Russell, \PR(D73,085002,2006); \\  
Erratum, {\it ibid.} {\bf D74}, 019902 (2006).

\bibitem{UED5n2}
M.~Battaglia, et al, JHEP {\bf 0507}, 033 (2005); \\
A.~Datta, K.~Kong and K.~T.~Matchev, Phys.\ Rev.\  D {\bf 72}, 096006 (2005) [Erratum-ibid.\  D {\bf 72}, 119901 (2005)].

\bibitem{LHTspectrum}
I. Low, JHEP {\bf 10}, 067 (2004).

\bibitem{SM4lowenergy}
M.~S.~Chanowitz, Phys.\ Rev.\  D {\bf 79}, 113008 (2009).

\bibitem{SM4bounds}
Review of Particle Physics, C. Amsler et al, Phys. Lett. B, {\bf 667}, 1 (2008).

\bibitem{SUSYreview}
A fairly comprehensive list of references can be found at the URL \\
http://lhcsigs.physics.lsa.umich.edu/mediawiki/index.php/Signature\_Reference

\bibitem{UED6}
B.A.~Dobrescu and E.~Ponton, JHEP {\bf 0403}, 071 (2004); \\
G.~Burdman, B.A.~Dobrescu and E.~Ponton, JHEP {\bf 0602}, 033 (2006);\\ 
K.~Ghosh, A.~Datta, Phys. \ Lett.\ B \ {\bf 665}, 369 (2008). 

\bibitem{suspect}
A.~Djouadi, J.~L.~Kneur and G.~Moultaka, Comput.\ Phys.\ Commun.\  {\bf 176}, 426 (2007).

\bibitem{LHTlowenergy}
X.~F.~Han, arXiv:0908.2572 [hep-ph].

\bibitem{SUSYbounds}
A.~Djouadi, M.~Drees and J.~L.~Kneur, JHEP {\bf 0603}, 033 (2006)

\bibitem{LesHouches}
J.~Alwall et al, Comput.\ Phys.\ Commun.\  {\bf 176}, 300 (2007).

\bibitem{LHTbounds}
J.~Hubisz and P.~Meade, Phys. \ Rev.\ D \ {\bf 71}, 035016 (2005). 

\bibitem{Belyaev}
See the URL \\
http://www.hep.phys.soton.ac.uk/~belyaev/public/calchep/index.html

\end{thebibliography}
\end{document}